\documentclass[aps,prd,reprint,nofootinbib, preprintnumbers]{revtex4-1}

\usepackage{bm}
\usepackage{graphicx} %
\usepackage{amsmath} %
\usepackage{amssymb} %
\usepackage{url}
\usepackage{subfigure} %
\usepackage{float} %
\usepackage{upgreek} %
\usepackage{multirow} %
\usepackage{color} %
\usepackage{lineno} %
\usepackage{hyperref} %
\usepackage{booktabs}
\usepackage{placeins}
\usepackage{ulem}
\usepackage[usenames,dvipsnames]{xcolor}
\hypersetup{colorlinks=true, urlcolor=black, linkcolor=blue, citecolor=red} %

\def\LCDM{$\Lambda$\rm{CDM}}
\def\ihMpc{h{\rm Mpc}^{-1}}
\def\hMpc{h^{-1}{\rm Mpc}}

\usepackage{amsmath}
\usepackage{tcolorbox}
\allowdisplaybreaks[4]

\begin{document}

\title{Lagrangian displacement field estimators in cosmology}

\author{Atsuhisa Ota${}^{1,2}$}
\email{iasota@ust.hk}
\author{Hee-Jong Seo${}^2$}
\author{Shun Saito${}^{3,4}$}
\author{Florian Beutler${}^{5}$}
\affiliation{${}^1$HKUST Jockey club Institute for advanced study, Hong Kong University of science and technology, Clear Water Bay, HK, PRC}
\affiliation{${}^2$Department of Physics and Astronomy, Ohio University, Athens, OH, 45701, USA}
\affiliation{${}^{3}$ Institute for Multi-messenger Astrophysics and Cosmology, Department of Physics\\
Missouri University of Science and Technology, 
1315 N. Pine St., Rolla MO 65409, USA}
\affiliation{${}^{4}$Kavli Institute for the Physics and Mathematics of the Universe (WPI), Todai Institutes for Advanced Study,\\
the University of Tokyo, Kashiwanoha, Kashiwa, Chiba 277-8583, Japan}
\affiliation{${}^{5}$Institute for Astronomy, University of Edinburgh, Royal Observatory, Blackford Hill, Edinburgh EH9 3HJ, UK}

\date{\today}

\begin{abstract}

The late-time nonlinear Lagrangian displacement field is highly correlated with the initial field, so reconstructing it could enable us to extract primordial cosmological information. 
Our previous work~\cite{Ota:2021caz} carefully studied the displacement field reconstructed from the late time density field using the iterative method proposed by Ref.~\cite{Schmittfull:2017uhh} and found that it does not fully converge to the true, underlying displacement field (e.g., $\sim 8\%$ offset at $k\sim 0.2 \ihMpc$ at $z=0.6$). We also constructed the Lagrangian perturbation theory model for the reconstructed field, but the model could not explain the discrepancy between the true and the reconstructed fields in the previous work. The main sources of the discrepancy were speculated to be a numerical artifact in the displacement estimator due to the discreteness of the sample. In this paper, we develop two new estimators of the displacement fields to reduce such numerical discreteness effect, the normalized momentum estimator~(NME) and the rescaled resumed estimator~(RRE). We show that the discrepancy Ref.~\cite{Ota:2021caz} reported is not due to the numerical artifacts. We conclude that the method from Ref.~\cite{Schmittfull:2017uhh} cannot fully reconstruct the shape of the nonlinear displacement field at the redshift we studied, while it is still an efficient BAO reconstruction method. In parallel, by properly accounting for the UV-sensitive term in a reconstruction procedure with an effective field theory approach, we improve the theoretical model for the reconstructed displacement field, by almost five times, from $\sim 15\%$ to the level of a few \% at $k\sim 0.2\ihMpc$ at the redshift $z=0.6$.

 

\end{abstract}

\maketitle


\section{Introduction}

Baryon acoustic oscillation~(BAO) imprints the sound horizon scale at recombination, which can be used to infer information about the nature of dark energy. The resulting precision critically depends on the strength of the BAO signal while the signal has been smeared as matter travels from the initial locations during nonlinear structure formation~\cite[e.g.,][]{Eisenstein:2006nj, Crocce:2007dt, Seo:2009fp}.  
However, the displacement of each mass tracer is mostly free from the degradation effect~\cite{Baldauf:2015tla}, as theoretically suggested by the Lagrangian resummation theory, where the exponential damping factor appears after resuming the 1-loop density power spectrum by using the Lagrangian displacement~\cite{Matsubara:2007wj}. 
Therefore, the Lagrangian displacement could be a useful degradation-free alternative, for extracting the BAO information, to the traditional observables based on the Eulerian fluid dynamics~\cite{Schmittfull:2017uhh}.

\medskip
Estimating the true displacement field from observed mass tracers is not straightforward for real surveys. First, we only measure the final locations of mass tracers. Also, there is a technical difficulty in estimating an unbiased displacement field from discrete, subsampled tracers as we will incorrectly measure the vanishing displacement field at the location where tracers do not exist.

Recently, there have been promising extensions of the standard density field reconstruction \citep{Eisenstein:2006nk} suggested by various groups \cite[e.g.,][]{Tassev:2012hu,Wang_2017, Zhu:2016xyy,Zhu:2016sjc,Yu_2017,Schmittfull:2017uhh,Shi_2018,Hada:2018fde,Hada:2018ziy,MaoCNN_2021,RecCNN:2022}. Among these, Ref. \cite{Schmittfull:2017uhh} is one of the subset (for example, the method by Refs.~\cite{Zhu:2016sjc,Zhu:2016xyy,Yu_2017,Shi_2018} is also designed to derive the displacement field) that more directly focuses on the aspect of reconstructing the displacement field.

The method was demonstrated to return a superior BAO reconstruction performance, compared to the standard method \cite{Eisenstein:2006nk}, particularly at a very low shot noise regime \cite{Ota:2021caz, Seo:2021nev}. In detail, this method attempts to find the uniform Lagrangian frame by displacing each observed galaxy particle along the local density gradient, progressively reducing the smoothing scale. Once we achieve the almost homogenous mass distribution, we can estimate the displacement field by measuring the difference between the Eulerian and the estimated Lagrangian positions. If we can indeed recover the true nonlinear displacement field from such a method, the broadband shape of the resulting clustering could be modeled utilizing the perturbation theories of the displacement field~\cite[e.g.,][]{Baldauf:2015tla}, allowing a cosmological parameter extraction from the shape of the power spectrum in addition to the reconstructed BAO feature.

In Ref.~\cite{Ota:2021caz}, we constructed a theoretical model for the displacement reconstruction by Ref.~\cite{Schmittfull:2017uhh}; there, we found that, while the method can reconstruct the BAO very well, it does not recover the shape of the true displacement field, and our theoretical model could not explain the deviation. Sources of the discrepancy were speculated, particularly including a numerical artifact due to the discreteness of the tracer sample for estimating the vector field.

This paper develops two new estimators of displacement fields that may reduce such discreteness effects in the tracers. The first estimator removes the effect of the tracer distribution, utilizing the fact that the Lagrangian positions of the (subsampled) particles are uncorrelated with the nonlinear displacement field. The second estimator asymptotically converts the displacement field to a density field.

We first test these estimators for measuring the true displacements from simulations (i.e., knowing the initial locations of the particles exactly) for various subsampling levels. While we perform this test as a sanity check before applying them to the reconstructed field, this application could be useful in estimating the nonlinear displacement field with sampling noise.

We then apply the estimators to the reconstructed displacement field and show that the discrepancy between the reconstructed and the true displacement fields remains even at $k~\sim 0.1\ihMpc$ and it is physical, not due to a numerical artifact. We, therefore, show that the method from Ref.~\cite{Schmittfull:2017uhh} cannot fully recover the true nonlinear displacement field near the redshift we studied ($z=0.6$), while it is still an effective BAO reconstruction method. 
By better securing the results against the discreteness effect, we improve the theoretical model for the reconstructed displacement field by properly accounting for the UV-sensitive term in a reconstruction procedure with an effective field theory approach.


\medskip
We organize this paper as follows.
In section~\ref{secNME}, we review the discreteness effect in galaxy surveys and pose a question about the displacement field estimators.
Sections~\ref{NMEsec} and \ref{rescale} introduce two new displacement field estimators and give a theoretical background for them.
A comparison of those new estimators and the previous mass-weighted one is presented in Section~\ref{sims}.
Then we apply the estimators for the iterative reconstruction in Section~\ref{secPost}, and we discuss the remaining inconsistencies in the post-reconstruction estimators, the true displacement field, and the 1-loop perturbation theory modeling.
The final section is devoted to the conclusions.

\section{Discreteness effect in galaxy surveys}\label{secNME}

This section investigates the theoretical aspects of the discreteness effect in galaxy surveys.
We first briefly review the mathematics for the discreteness effect in the galaxy field.
Then, we illustrate the issue in the displacement field measurement.
This is a problem with interpreting a vector field from measuring mass tracers.

\subsection{Galaxy number density field}\label{Po:G}

Let $\bar n_g$ be the average galaxy number density and $V$ be the volume of a given three-dimensional pixel.
Then the probability that the number of galaxies $N$ found in the pixel follows a Poisson distribution whose average is $\bar n_g V$, i.e.,  
\begin{align}
	 e^{-\bar n_g V}\frac{(\bar n_g V)^{N}}{N!}.\label{pod}
\end{align}
Using the Poissonian random variable at the position $\mathbf x$, i.e., $N(\mathbf x)$, the galaxy number density field is given as
\begin{align}
	n_g(\mathbf x) = \frac{N(\mathbf x)}{V(\mathbf x)}.\label{np}
\end{align}
The 2-point correlation function of the number density is given by
\begin{align}
	\langle n_g(\mathbf x) n_g(\mathbf y) \rangle_{\rm Po} = \left \langle \frac{N(\mathbf x)}{V(\mathbf x)} \frac{N(\mathbf y)}{V(\mathbf y)} \right \rangle_{\rm Po}, 
\end{align}
where the subscript ``Po'' means that the average is taken by the locally defined Poisson distribution of Eq.~\eqref{pod}.
For $\mathbf x= \mathbf y$, we get 
\begin{align}
	\langle n_g(\mathbf x)^2 \rangle_{\rm Po} = \frac{\bar n_g}{V(\mathbf x)}. 		
\end{align}
For $\mathbf x \neq \mathbf y$, the distributions are uncorrelated, and we find
\begin{align}
	\langle n_g(\mathbf x)n_g(\mathbf y) \rangle_{\rm Po}=\langle n_g(\mathbf x)\rangle_{\rm Po} \langle n_g(\mathbf y) \rangle_{\rm Po} = \bar n_g^2.
\end{align} 
To summarize, we derive~\cite{1980lssu.book.....P}
\begin{align}
	\langle n_g(\mathbf x)n_g(\mathbf y) \rangle_{\rm Po} = \frac{\delta_{\mathbf x,\mathbf y}}{V(\mathbf x)}\left(
	\bar n_g - V(\mathbf x)\bar n_g^2
	\right) + \bar n_g^2, \label{14}
\end{align}
where $\delta_{\mathbf x,\mathbf y}=1$ for $\mathbf x=\mathbf y$, and otherwise zero.
In the small pixel limit~($V(\mathbf x)\bar n_g \ll 1$), defining the Poisson noise $\delta_g \equiv (n_{g}-\bar n_{g})/{\bar n_{g}}$, we get
\begin{align}
	\langle \delta_g(\mathbf x)\delta_g(\mathbf y) \rangle_{\rm Po}  = \frac{1}{\bar n_g} \delta^{(3)}_{\rm D}(\mathbf x-\mathbf y),
\end{align}
where $\delta_{\mathbf x,\mathbf y}/V(\mathbf x) \approx \delta^{(3)}_{\rm D}(\mathbf x-\mathbf y)$ is the three-dimensional Dirac's delta function.
For simplicity, we assume that the galaxy distribution is linearly related to the underlying dark matter density fluctuation $\delta_m$.
Then the number fluctuation due to the primordial density field is written as $b \delta_m$.
We replace $\bar n_g$ with the local number $\bar n_g (1+b\delta_m)$ when we normalize the Poisson distribution of Eq.~\eqref{pod}, and we derive 
\begin{align}
	\langle \delta_g(\mathbf x)\delta_g(\mathbf y) \rangle_{\rm Po,G} = 	
	\frac{\delta^{(3)}_D(\mathbf x-\mathbf y)}{\bar n_g} 
	 + b^2\langle \delta_m(\mathbf x) \delta_m(\mathbf y)\rangle_{\rm G},\label{shot_gx}
\end{align}
where ``G'' implies the Gaussian average of the primordial density perturbations.
The first term comes from the Poisson shot noise, while the second term is the cosmological signal.

\subsection{Displacement field}

Next, we consider the discreteness effect for the displacement field.
For a given Eulerian coordinate $\mathbf x$ and its associated Lagrangian coordinate $\mathbf q$, the displacement field is defined as
\begin{align}
	\mathbf x = \mathbf q + \mathbf \Psi(\mathbf q).\label{dispdef}
\end{align}
This $\mathbf \Psi(\mathbf q)$ in Eq.~\eqref{dispdef} is a continuous field, and in practice, we may interpret the displacement field by measuring the position of a test particle on the field. We assign the individual particle displacements to the grid in data analysis. 
Let $V(\mathbf q)$ be a pixel volume at $\mathbf q$, and $N(\mathbf q)$ be the number of particles found in the pixel, which is a Poissonian variable defined at a Lagrangian position $\mathbf q$.
One may evaluate the center of the mass displacement field 
\begin{align}
	\tilde{\mathbf \Psi}(\mathbf q) = \frac{\sum_{i=1}^{N(\mathbf q)} \mathbf \Psi(\mathbf q_i)  }{N(\mathbf q)}\label{25},
\end{align}
for the cell at $\mathbf q$ where $i$ is the particle label in the pixel.
The issue is that Eq.~\eqref{25} is ill-defined for empty pixels, i.e., for $N(\mathbf q)=0$. 
One may consider interpolating the value at the empty pixels, e.g., based on the values at the neighboring pixels, but any ad hoc prescription to $N$ = 0 will introduce additional complexity in estimating the effect of the Poisson fluctuations on the displacement estimator.
More importantly, the interpolation becomes quickly inefficient because most pixels (assuming a reasonable pixel resolution, e.g., 5{\rm Mpc/h} for $\bar n_{g}$=0.001 ${\rm h}^3/{\rm Mpc}^3$)  would be empty in a sparse system such as the galaxy field.
In Refs.~\cite{Ota:2021caz, Seo:2021nev}, we selectively set $\tilde{\mathbf \Psi}=0$ for the empty pixel and attempt to correct the resulting large-scale effect by rescaling the clustering amplitude by a constant factor.
However, we indeed found that there still is a residual discrepancy in the small-scale power that depends on the sampling fraction (e.g., at the level of 6-7\% at $k\sim 0.3\ihMpc$ in Figure 2 between L500 and subL500). Thus, it was nontrivial to extract the unbiased displacement ﬁelds on small scales using the mass-weighted estimator (MWEs) we adopted.

\medskip
The MWE fails because the displacement field is irrespective of the occupation of the pixel.
An empty pixel does not mean the displacement of the pixel is zero.
Eulerian velocity estimators also suffer from a similar issue to the mass-weighted estimators.
To our knowledge, Ref.~\cite{Bernardeau:1995en} was the first to point out that mass-weighted velocity estimators return a biased estimator of the actual velocity ﬁeld since the measurements relying on the mass tracer counting give the momentum rather than the velocity.
They proposed two volume-weighted assignments, the Voronoi tessellation method and the Delaunay tessellation method for velocity measurements, which were applied for describing shell crossing in Ref.~\cite{Pueblas:2008uv}. 
Furthermore, various volume weighted assignments are proposed~\cite{Colombi:2008qz,Zheng:2013ora,Zhang:2014hra,Zheng:2014ywa,Yu:2015gla,Yu:2016mzj}.
On the other hand, giving up the displacement/velocity and considering the momentum field can also be an option~\cite{Park:2000rc,Park:2005bu,Park:2000rc,Howlett:2019bky,Pan:2020thr}.
An advantage of the momentum field is that the sampling issue is solved as momentum is, correctly, zero without a mass tracer. However, higher-order effects such as the galaxy bias would complicate estimating the true momentum field.
While the displacement field discreteness effects are similar to the velocity, we focus on the properties specific to the Lagrangian perspective to find a solution to our issue. The following sections introduce new displacement estimators that can reduce the discreteness effect.

\medskip
As a caveat, while we first test our new estimators with the true displacement fields in a simulation, we note that the displacement field is not directly observable in real surveys since we only measure the final Eulerian position of each galaxy.
Therefore, we are interested in ``reconstructing'' the Lagrangian position from the observed Eulerian position, i.e., the displacement field. Our new estimators are developed to interpret the reconstructed displacement field properly. The reconstruction scheme to find the Lagrangian frame itself was also discussed in Refs.~\cite{Schmittfull:2017uhh, Ota:2021caz, Seo:2021nev}, and we review the idea in Section~\ref{secPost}.

\section{Normalized Momentum Estimator}
\label{NMEsec}
This section introduces a new momentum estimator and shows how one can normalize it to obtain a volume-weighted displacement field estimator to a good approximation.

\subsection{Definition}
We propose to compute a normalized momentum estimator~(NME)
\begin{align}
		\tilde \xi_{ij}(\mathbf q-\mathbf r) &\equiv \frac{\left\langle \sum_{a=1}^{N(\mathbf q)}\sum_{b=1}^{N(\mathbf r)}\Psi_i(\mathbf q_a) \Psi_j(\mathbf r_b)\right \rangle}{\left \langle N(\mathbf q)N(\mathbf r)\right\rangle} 
		,\label{tildexi}
\end{align}
and we will show that $\tilde \xi_{ij} \to \xi_{ij}\equiv \langle \Psi_i(\mathbf q) \Psi_j(\mathbf r)\rangle$ for the small pixel limit below.
With this approximation, we can avoid the ill-defined mass-weighted displacement in Eq.~\eqref{25}.
In Eq.~\eqref{tildexi}, one first computes the numerator, which is the correlation function of a momentum-like quantity, and then we normalize the correlation function by the density correlation function.
This would allow us to construct an approximately volume-weighted estimator without, e.g., the Delaunay tessellation method.

\begin{figure}
	  \centering \includegraphics[width=0.6\linewidth]{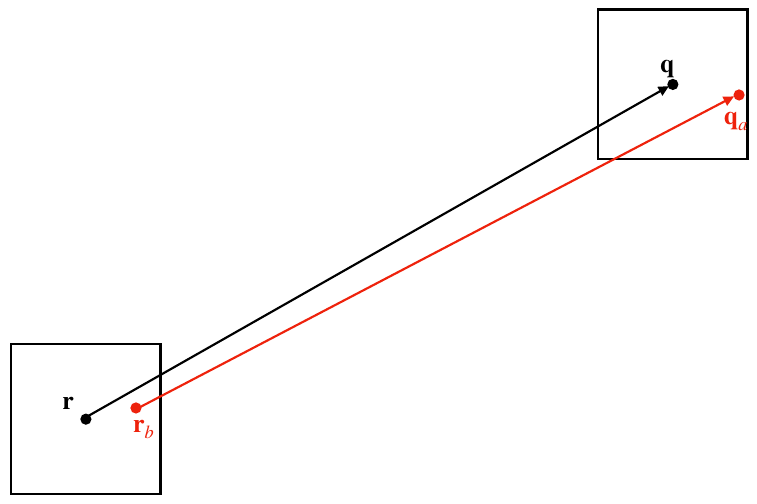}
	  \caption{An example configuration for $\mathbf q$, $\mathbf q_a$, $\mathbf r$ and $\mathbf r_b$. $\epsilon_{ab} \equiv |\mathbf q_a- \mathbf r_b| - |\mathbf q- \mathbf r|$ is within the size of a box.
	  }
	  \label{fig_pf}
\end{figure}

\medskip
Let us prove $\tilde \xi_{ij} \to \xi_{ij}$ for $V\to 0$ limit.
For simplicity, the particles are Poisson sampled in the uniform Lagrangian frame. 
The bracket of Eq.~\eqref{tildexi} means that we take both the Poisson and ensemble average.
$N(\mathbf q)$ and $N(\mathbf r)$ in Eq.~\eqref{tildexi} are independent of the ensemble average in Lagrangian space 
so that we may exclusively take the ensemble average for the displacement field. $\xi$ is a function of the distance due to the statistical isotropy and homogeneity.
Then Eq.~\eqref{tildexi} is expanded into
\begin{align}
	&\tilde \xi_{ij}(\mathbf q-\mathbf r)	= \left.\sum_{n=0}^{\infty} \mathcal F_{n}(\mathbf q,\mathbf r)\partial_x^n\xi_{ij}(x)\right |_{x=|\mathbf q-\mathbf r|},\label{eq35}
\end{align}
where we used the Taylor expansion 
\begin{align}
	\xi_{ij}(\mathbf q_a -\mathbf r_b) &=\left.\sum_{n=0}^{\infty} \frac{\epsilon_{ab}^n}{n!}\partial_x^n\xi_{ij}(x)\right |_{x=|\mathbf q-\mathbf r|} ,\\
	\epsilon_{ab} &\equiv |\mathbf q_a  - \mathbf r_b|-|\mathbf q - \mathbf r|,
\end{align}
and then defined
\begin{align}
	\mathcal F_n(\mathbf q,\mathbf r) &\equiv  \left\langle \sum_{a=1}^{N(\mathbf q)}\sum_{b=1}^{N(\mathbf r)}\right \rangle_{\rm Po}^{-1}   \left\langle \sum_{a=1}^{N(\mathbf q)}\sum_{b=1}^{N(\mathbf r)} \frac{\epsilon_{ab}^n}{n!} \right \rangle_{\rm Po}.
\end{align}
We cannot explicitly evaluate $\mathcal F_n$ because we never know the implicit Poissonian dependence of $\epsilon_{ab}$, i.e., the Poissonian dependence of the particle positions. However, we can find the upper bound easily.
An example configuration for $\mathbf q$, $\mathbf q_a$, $\mathbf r$ and $\mathbf r_b$ is illustrated in Fig.~\ref{fig_pf}.
Let $V(\mathbf q)$ be a cube and $L=V^{\frac13}$.
The maximum distance between particles in the same box is $\sqrt{3}L$, so we get 
\begin{align}
	|\mathcal F_n |\leq \frac{(\sqrt{3}L)^n}{n!}.\label{352}
\end{align}
Then one finds
\begin{align}
 &\left| \tilde \xi_{ij}(|\mathbf q-\mathbf r|) -  \xi_{ij}(|\mathbf q-\mathbf r|) \right|\notag \\
 &\leq \sum_{n=1}^{\infty} \frac{(\sqrt{3}L)^n}{n!}\left | \partial_x^n\xi_{ij}(x)\right |_{x=|\mathbf q-\mathbf r|}.
\end{align}
We can take a sufficiently small $V$ such that $L^n \partial^n\xi_{ij}\to 0$, as long as $\partial^n \xi_{ij}$ is finite.

\medskip
The above proof is applicable as particle sampling is independent of the primordial fluctuations, so the distribution of particles does not have to be Poissonian as long as it is independent of the ensemble average.
Interestingly, the mass window is canceled in Eq.~\eqref{tildexi} as we have $\mathcal F_0=1$ so that we obtain the volume-weighted correlation function, the shot noise is zero, and the final result is independent of the number density $\bar n_{\rm g}$.
We would not claim that the covariance vanishes, but the discreteness effect vanishes.
The above proof is not valid for a singular $\xi$.

Eq.~\eqref{tildexi} is different from the nominal momentum estimator in Ref.~\cite{Pan:2020thr} that corresponds to
\begin{align}
    \left\langle \sum_{a=1}^{N(\mathbf q)}\sum_{b=1}^{N(\mathbf r)}\Psi_i(\mathbf q_a) \Psi_j(\mathbf r_b)\right \rangle.\label{MOn}
\end{align}
Eq.~\eqref{MOn} is a mass-weighted quantity without the normalization in the denominator of Eq.~\eqref{tildexi} and therefore does not correspond to a volume-weighted estimator.

In Eulerian space, e.g., for the late time velocity field, we cannot exclusively take the ensemble average in Eq.~\eqref{tildexi}. This is because the Poisson distribution depends on the local stochastic variable, as discussed in Sec.~\ref{Po:G}. Hence, the bispectrum or higher-order cumulant appears and cannot be canceled. Therefore, the above proof only applies to the Eulerian velocity field at leading order perturbations.

Another crucial remark is that because the new estimator is effectively volume-weighted, we expect that Eq.~\eqref{tildexi} is a promising estimator for very sparse, such as a biased galaxy tracer. In our future work, we plan to investigate how well such an estimator of the galaxy field would relate to the matter displacement field as a function of scale.

\subsection{Numerical implementation}
\def \bx{\mathbf x}
\def \bk{\mathbf k}
Below, we compare the mass-weighted estimator used in Ref.~\cite{Ota:2021caz} with the operation conducted for the NME in this paper.
 
The former mass-weighted displacement was defined as
\begin{align}
	\mathbf \Psi^{\rm obs.}_p 
	&= \frac{\sum_i W_{\rm CIC}({\bx_p,\bx_i}) \mathbf \Psi^{\rm obs.}(\bx_i)  }{\sum_i W_{\rm CIC}({\bx_p,\bx_i}) }
	,\label{mweight}
\end{align}
which corresponds to Eq.~\eqref{25}.
Here $\mathbf \Psi^{\rm obs}(\bx_i)$ can be either the true displacement or the iteratively reconstructed displacement of each mass tracer labeled by $\mathbf x_i$.
$W_{\rm CIC}$ is the pixel window function indicating that
we are using the cloud-in-cell assignment, and $\mathbf \Psi^{\rm obs.}_p$ is the estimator assigned at a pixel centered at $\bx_p$.

For NME, we have the same catalog of $\mathbf \Psi^{\rm obs}(\bx_i)$, but we directly evaluate its power spectrum without explicitly introducing the displacement field estimators like Eq.~\eqref{mweight}.
First, we estimate the momentum field 
\begin{align}
   \mathbf P^{\rm obs.}_p = \sum_i W_{\rm CIC}({\bx_p,\bx_i}) \mathbf \Psi^{\rm obs.}(\bx_i) \, .
\end{align}
Then, we consider fast Fourier transformation~(FFT) of the momentum field $\hat{\mathbf P}^{\rm obs.}_p$ and evaluate the momentum power spectrum $\langle \hat{\mathbf P}^{\rm obs.}_p(\mathbf k_1) \cdot \hat{\mathbf P}^{\rm obs.}_p(\mathbf k_2)\rangle$.
Then, we compute the correlation function of the momentum field $\xi_P$ from the momentum power spectrum.
We also calculate the correlation function $\xi_n$ of the number density field (i.e., the discreteness selection function):
\begin{align}
   n^{\rm obs.}_p = \sum_i W_{\rm CIC}({\bx_p,\bx_i}).
\end{align}
Then, we compute $\xi_P/\xi_n$ to convert the momentum field two-point statistics to the displacement field statistics, and finally, Fourier transform it to find the displacement field divergence spectrum.
We ignored the curl component for simplicity so we have $(\mathbf k\cdot \mathbf \Psi)^2 = k^2 \mathbf \Psi\cdot \mathbf \Psi$. As emphasized in the previous section, because $\bx_i$ is in the Lagrangian space for the true displacement field or presumably close to the true Lagrangian space in the reconstructed case, we can ignore the higher order cumulants in $\xi_n$ and $\xi_P$. Again, the same operation would not be justified in calculating a late time field where $\bx_i$ is not close to the Lagrangian space, e.g., the Eulerian velocity field.
To compare the density power spectrum, we use $k^2P_{\mathbf \Psi}$ rather than $P_{\mathbf \Psi}$.
In section~\ref{sims}, we compare the NME and the exact displacement field.

\section{Rescaled resummation estimator} \label{rescale}

This section describes another displacement field estimator for a sparse sample we propose in this paper, which we call the rescaled resummation estimator~(RRE).
We measure the displacement field divergence power spectrum as a rescaled density power spectrum without explicitly evaluating the vector field on the grid; therefore, we can avoid the issue of displacement field assignment with sparse tracers.

\subsection{Lagrangian resummation thoery}

Let us consider the density fluctuation exponentiated by the rescaled displacement field $\mathbf \Psi_{\rm NL}/\Lambda$:
\begin{align}
	\delta^{\Lambda}_{\rm NL} &\equiv \int d^3 q e^{-i\mathbf k \cdot \mathbf q} \left(e^{-i\mathbf k \cdot   \frac{\mathbf \Psi_{\rm NL}(\mathbf q)}{\Lambda} }-1 \right).\label{deltanrec}
\end{align}
For $\Lambda=1$, this equation recovers a known integral representation of the density perturbation in Lagrangian perturbation theory.
In simulations, Eq.~\eqref{deltanrec} corresponds to measure the density fluctuation after moving the particles from their initial/Lagrangian locations by $\mathbf \Psi_{\rm NL}/\Lambda$.
Taylor expanding Eq.~\eqref{deltanrec}, we get
\begin{align}
    \Lambda \delta^{\Lambda}_{\rm NL}  = i\mathbf k\cdot \mathbf \Psi_{\rm NL} + \mathcal O ( \Lambda^{-1}),\label{19mult}
\end{align}
and we get the exact nonlinear displacement field in the $\Lambda \to \infty$ limit.
From Lagrangian resummation theory~\cite{Matsubara:2007wj}, the resummed power spectrum of Eq.~\eqref{deltanrec} is given as
\begin{align}
	\Lambda ^2 P_{\delta^{\Lambda}_{\rm NL}} =& \exp\left(-\frac{k^2\int p^2dp P_{\Psi_{\rm NL} }}{6\pi^2 \Lambda^2} + \mathcal O(\Lambda^{-3}) \right) 
	\notag \\
	&\times \left ( k^2 P_{\Psi_{\rm NL}} + \mathcal O(\Lambda^{-1})\right).\label{63}
\end{align}
The correction terms are higher-order cumulants in $\Psi_{\rm NL}$, which are relatively suppressed to the leading order power spectrum since they carry additional negative powers of $\Lambda$.
The exponential damping is the degradation effect due to the dark matter displacement from the initial BAO configurations.
After rescaling \eqref{deltanrec}, the damping effect is reduced thanks to the $\Lambda^2$ in the denominator of the exponential and the estimator asymptotes to the displacement field power spectrum. We obtain the displacement field divergence power spectrum without explicitly evaluating vector fields and the power spectrum as the density power spectrum by taking a large $\Lambda$ limit.%

\subsection{Noise modeling}
A large $\Lambda$ will improve the recovery of the displacement field power spectrum (and the BAO feature in it), but
multiplying large $\Lambda$ may also amplify the noise in the density field.
This subsection illustrates the potential issue and how to mitigate some effects.
A possible error comes from the uncertainty $\bm \chi$ in the estimated Lagrangian position during reconstruction, i.e., the estimated Lagrangian position should be written as $\mathbf q + {\bm \chi}$ rather than $\mathbf q$.
Due to this error, the estimated displacement field is also shifted by the same amount.
Therefore, the rescaled particle location will be replaced as 
\begin{align}
     \mathbf q + \frac{\mathbf \Psi_{\rm NL}}{\Lambda} \to  \mathbf q+{\bm \chi} +\frac{\mathbf \Psi_{\rm NL} - {\bm \chi} }{\Lambda}.\label{noisemod}
\end{align}
We cannot isolate ${\bm \chi}$ from the observed data, so the rescaling happens only for $\mathbf \Psi_{\rm NL} - {\bm \chi}$.
Nonvanishing ${\bm \chi}$ is generally inevitable when reconstructing the displacement field from actual data. 
Eq.~\eqref{deltanrec} is generalized to
\begin{align}
	\delta^{\Lambda}_{\rm NL}\to  \int d^3 q e^{-i\mathbf k \cdot \mathbf q} \left[e^{-i\mathbf k \cdot \left( \frac{(\Lambda -1){\bm \chi}(\mathbf q)}{\Lambda} +   \frac{\mathbf \Psi_{\rm NL}(\mathbf q)}{\Lambda} \right)}-1 \right].
\end{align}
The $\chi$ term in the above equation is not scaled for a large $\Lambda$, so the noise term is relatively amplified after multiplying $\Lambda$ in Eq.~\eqref{19mult}. 
The resummed power spectrum is given as
\begin{align}
	&\Lambda^2 P_{\delta^{\Lambda}_{\rm NL}} 
	\to \exp\left[-\frac{k^2\int p^2dp }{6\pi^2 }\left(\frac{P_{\Psi_{\rm NL}}}{\Lambda^2} + \frac{(\Lambda-1)^2}{\Lambda^2}P_{\chi}\right)  \right] 
	\notag \\
	&\times k^2 \left [  P_{\Psi_{\rm NL}} + 2 (\Lambda -1) P_{\chi \Psi_{\rm NL}} + (\Lambda-1)^2 P_{\chi}  \right],\label{23}
\end{align}
where we ignored $\mathcal O(\chi^3,\Lambda^{-1})$ terms.
$\chi$ and $\Psi_{\rm NL}$ are generally correlated. 
Eq.~\eqref{23} implies that the RRE works well without amplifying noise when we have the following relation:
\begin{align}
    (\Lambda-1)^2\ll \frac{P_{\Psi_{\rm NL}}}{P_{\chi}},~\Lambda -1 \ll \frac{P_{\Psi_{\rm NL}}}{P_{\Psi_{\rm NL}\chi}},\label{katei}
\end{align}
which will not be satisfied for large $\Lambda$ or large inhomogeneities in $\chi$.
How can we then mitigate the amplification of the $\chi$ contribution?
We provide a simple method as follows.
By setting $\Lambda=\infty$ we can estimate the noise field as
\begin{align}
	\delta^{\infty}_{\rm NL} = \int d^3 q e^{-i\mathbf k \cdot \mathbf q} \left(e^{-i\mathbf k \cdot  {\bm \chi} }-1 \right) \label{36}.
\end{align}
This is the non-vanishing inhomogeneity in the estimated Lagrangian space, as reconstruction is not perfect. In our method, the error field can be approximately estimated from the final density field of the galaxies after a series of iterative reconstructions.
We can therefore subtract the noise field to define 
 \begin{align}
\hat \delta^{\Lambda}_{\rm NL} \equiv   \delta^{\Lambda}_{\rm NL} - \delta^{\infty}_{\rm NL}.\label{newest}
\end{align}
The corrected RRE is expanded into
\begin{align}
    \Lambda \hat \delta^{\Lambda}_{\rm NL} =i\mathbf k\cdot ( \mathbf \Psi_{\rm NL} - {\bm \chi})  + \mathcal O (\Lambda^{-1}).\label{lamdel}
\end{align}
Thus, the positive power of $\Lambda$ is removed so we can avoid amplifying the noise term, whereas the noise $\chi$ is inevitable.\footnote{We could have chosen to subtract by $(1-1/\Lambda)\delta^{\infty}_{\rm NL}$ in Eq~\eqref{newest} to cancel out the effect of $\chi$ more efficiently. We find that with that option, by boosting the small scale contribution slightly due to the factor $(1-1/\Lambda)$, the method reaches a better reconstruction at $n <6$  while the agreement with respect to the theory model and the NME is worse. For $n >8$ we find this choice also converges to NME.}
The resummed power spectrum has the following form
\begin{align}
	\Lambda^2 &P_{\hat \delta^{\Lambda}_{\rm NL}} = \exp\left[-\frac{k^2\int p^2dp }{6\pi^2 }P_{\chi} \right]  \notag \\
	&\times k^2 \left [  P_{\Psi_{\rm NL}} - 2 P_{\chi \Psi_{\rm NL}} +  P_{\chi} + \mathcal O(\chi, \Lambda^{-1})\right].\label{deltahatpower}
\end{align}
Now, the condition that $\Lambda \hat \delta^{\Lambda}_{\rm NL}$ converges to the displacement is independent from $\Lambda$, and Eq.~\eqref{katei} reduces to  
\begin{align}
    1 \ll \frac{P_{\Psi_{\rm NL}}}{P_{\chi}},~ 1 \ll \frac{P_{\Psi_{\rm NL}}}{P_{\Psi_{\rm NL}\chi}}.
\end{align}
The subtraction in Eq.~\eqref{newest} bears a resemblance to the operation in the standard BAO reconstruction estimator as well as the standard iterative scheme introduced in Ref.~\cite{Seo:2021}, which is written as the difference of the shifted reference density perturbation (denoted as $\delta_s$ in Ref.~\cite{Seo:2021}) and the displaced galaxy density perturbation ($\delta_d$ in Ref.~\cite{Seo:2021}). $\delta_d$ in standard/iterative reconstruction adds reconstructed small-scale information; likewise, we find that in RRE, subtracting with $\delta^{\infty}_{\rm NL}$ (i.e., $\delta_d$) recovers the small scale clustering by removing the most of the effect of $\chi$. We assume this procedure also subtracts the shot noise contribution to a good extent and therefore does not apply a separate shot-noise subtraction.

\section{New estimators in N-body simulations}\label{sims}

\begin{table*}
\caption{\label{table1}
Simulations and sampling parameters used in this paper. 
The simulations assume a flat $\Lambda$CDM cosmology in Ref.~\cite{Ade:2015xua}~($\Omega_{\rm m} = 0.3075$, $\Omega_{\rm b}h^2=0.0223$, $h=0.6774$, and $\sigma_8=0.8159$.).
}
\begin{ruledtabular}
\begin{tabular}{lccccc}
Name & subsampling \% & \# of meshes  & Box size [Mpc$/h]^3$ & \# of original particles &\# of simulations \\
\hline
fullL1500 &100  & $1536^3$  & $1500^3$ & $1536^3$ &1 \\
L500 &4 & $512^3$  & $500^3$ & $1536^3$ & 5\\
subL500 &0.15 & $512^3$  & $500^3$ & $1536^3$ & 5\\
subsubL500 & 0.015 & $512^3$  & $500^3$ & $1536^3$ & 5\\
\end{tabular}
\end{ruledtabular}
\end{table*}

We proposed the two new displacement field estimators in the previous sections and will now assess their performance in numerical simulations.
As mentioned earlier, we want to reconstruct the displacement field from actual galaxy surveys and apply the estimators for the post-reconstruction data.
In this section, we first test the methods by deriving the true displacement field (i.e., the difference between the initial Lagrangian positions and the final Eulerian positions in the simulations); we check if these estimators have advantages in mitigating the discreteness and/or subsampling effect.
We will then consider the post-reconstruction data in the next section.

This paper uses two different simulations from Ref.~\cite{Ota:2021caz} for different purposes. We briefly summarize the parameters of these simulations.
First, we focus on dark matter simulation rather than galaxies or halos for simplicity.
Both simulations are based on the flat \LCDM~cosmology in Ref.~\cite{Ade:2015xua} with $\Omega_{\rm m} = 0.3075$, $\Omega_{\rm b}h^2=0.0223$, $h=0.6774$, and $\sigma_8=0.8159$.
Full $N$-body simulations were produced using the MP-Gadget~\cite{Springel:2000yr,Springel:2005mi,2018zndo...1451799F} with the box size of $500{\rm Mpc}/h$ and $1500{\rm Mpc}/h$, and the simulations evolve $1536^3$ particles from $z=99$ by computing forces in a grid of $1536^3$.
For the former $500{\rm Mpc}/h$ simulation, we average five realizations and use only 4\% of the dark matter particles at $z=0.6$ for the data set named L500 and 0.15\% for subL500 in Tab.~\ref{table1}. 
We use a grid of $512^3$ to Fourier-transform and reconstruct this nonlinear field for L500 and subL500.
These data sets are prepared to test the robustness of the estimators: L500 has almost no empty FT grids, but the subsampled particles are not necessarily in the center of each mesh, while subL500 represents a sparse sample with 96\% of grids/meshes being empty.
We prepare the latter $1500{\rm Mpc}/h$ box simulation to generate the reference displacement field that we called the ``true displacement field''.
We name the set fullL1500, which samples all particles in the center of each grid in Lagrangian space.
Therefore, there is no empty pixel in the set, and the displacement measured in each grid is straightforwardly interpreted as the displacement field without sampling noise as a volume-weighted displacement.

In this work we compute the auto power spectrum of the displacement field divergence $P_{i\mathbf k\cdot {\bm \Psi}}$, and the cross-power spectrum of the displacement field divergence and the linear matter fluctuation, $P_{\delta_{\rm L}i\mathbf k\cdot {\bm \Psi} }$.
Both are normalized by the linear matter power spectrum $P_{\rm L}$.
We also refer $P_{\delta_{\rm L}\nabla \cdot {\bm \Psi} }/P_{\rm L}$ to the propagator $C_{i\mathbf k\cdot {\bm \Psi}} $.  
As we have only 1 or 5 realizations for simulations, it is difficult to derive reliable error bars. We therefore maximally utilize the variance cancellation by comparing propagators and power spectra relative to their corresponding initial conditions when presenting the results. 

\begin{figure}
	  \centering \includegraphics[width=\linewidth]{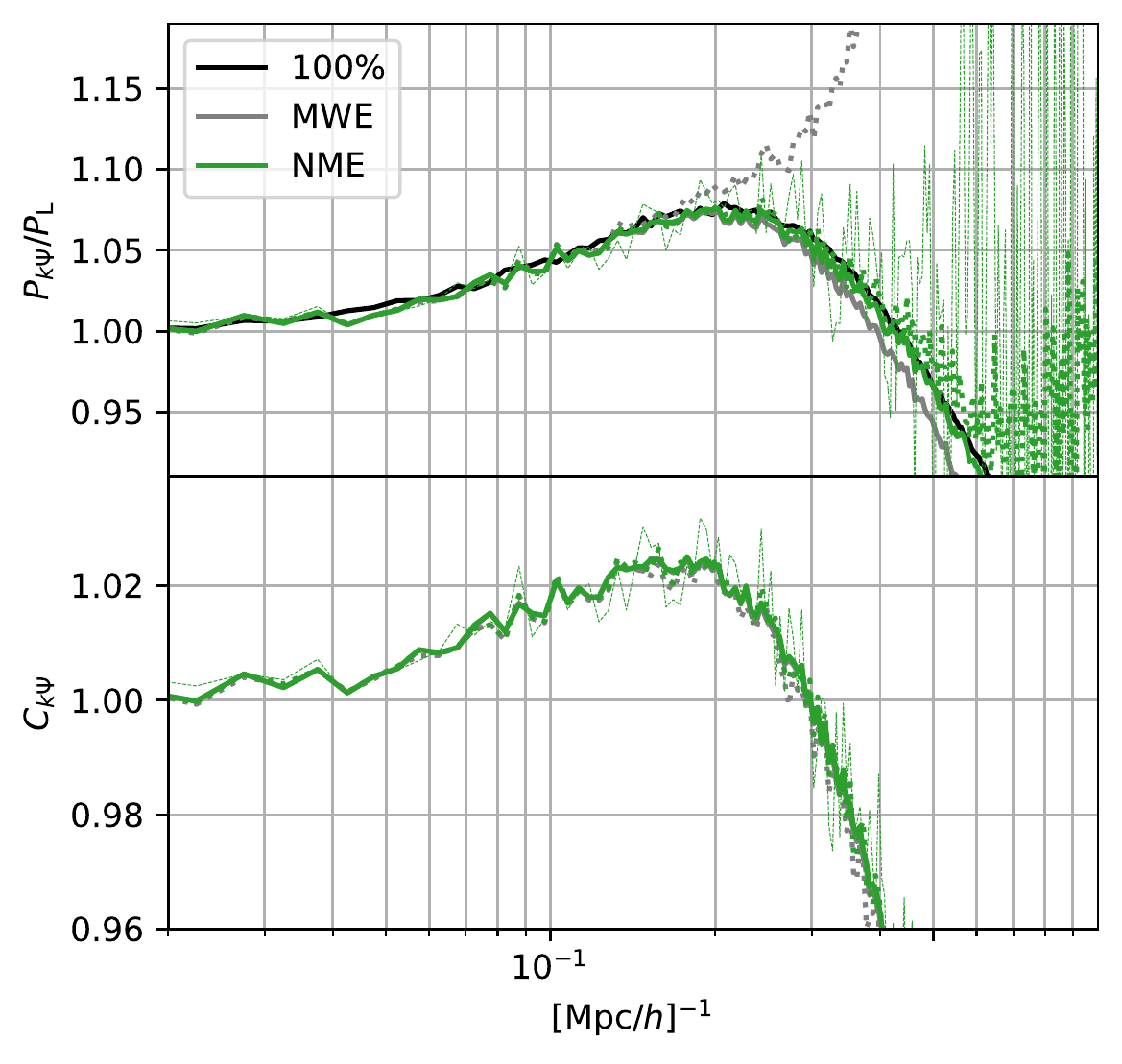}
  \caption{The displacement field divergence power spectra~(top) and propagators~(bottom) of the mass-weighted estimator~(MWE) and normalized momentum estimator~(NME) for various sampling fractions. The linear matter power spectrum normalizes the power spectrum, and the propagator is defined as the cross power spectrum normalized by the linear matter power spectrum. The figure is based on the N-body simulation L500 (solid) and subL500 (dotted) summarized in Tab.~\ref{table1}. 
   We also add subsubL500 for the NME~(faint dashed) that corresponds to $n_{\rm particle}=0.0043 h^3{\rm Mpc}^{-3}$, which still shows the convergence to the true displacement field in both panels despite the high sampling noise.
   The black solid curve corresponds to the fullL1500. 
  }
  	\label{figNME}
\end{figure}

\begin{figure}
	  \centering \includegraphics[width=\linewidth]{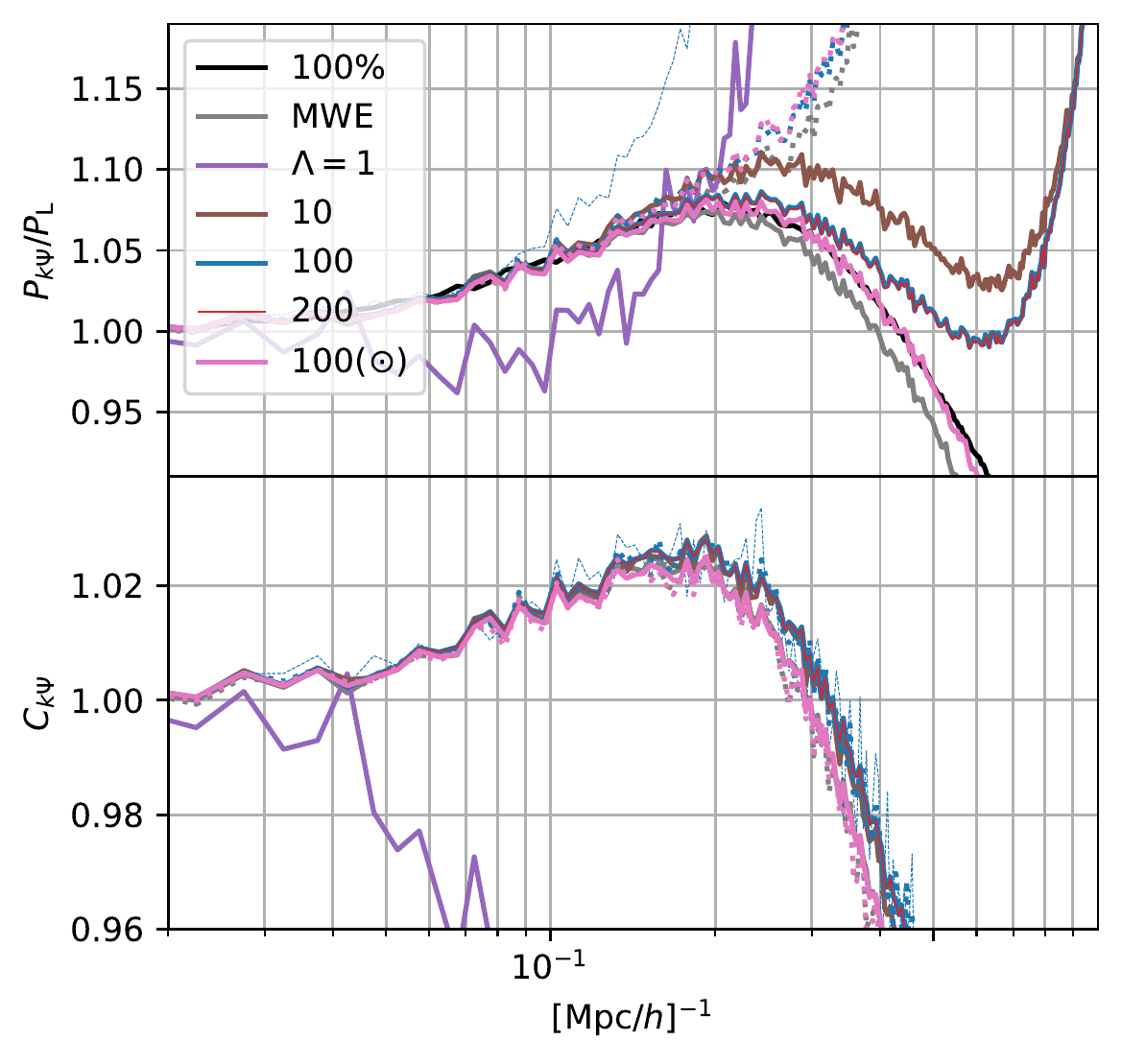}
  \caption{The displacement field divergence power spectra~(top) and propagators~(bottom) of the mass-weighted estimator~(MWE) and rescaled resummation estimator~(RRE) for various sampling fractions. The linear matter power spectrum normalizes the power spectrum, and the propagator is defined as the cross power spectrum normalized by the linear matter power spectrum. The figure is based on the N-body simulation L500 (solid) and subL500 (dotted) summarized in Tab.~\ref{table1}. 
The black solid curve corresponds to the fullL1500.
The violet line reproduces the nonlinear field with $\Lambda =1$.
  $\odot$ implies that we assigned the displacement fields at the center of each pixel.
  We again add subsubL500 for the RRE $\Lambda=100$ (faint blue dashed) that corresponds to $n_{\rm particle}=0.0043 h^3{\rm Mpc}^{-3}$. Contrary to NME, the top panel shows a residual contribution from the sampling noise, while the propagator fairly well converged to the true displacement field.}
 \label{figRRE}
\end{figure}

In Figs.~\ref{figNME} and~\ref{figRRE}, we show the displacement field power spectrum measured in L500, subL500, and fullL1500, normalized by the linear matter power spectrum for various estimators.
The NME and MWE converge for $k\lesssim 0.2\ihMpc$ within 1\%, but the MWE starts to depart from the true displacement field at higher $k$.
The NME (Figs.~\ref{figNME}) is more robust than the MWE.
A few percent convergences are extended to $k<0.5\ihMpc$; the solid green (L500) and dotted green (subL500) show much better consistency with respect to the true displacement field (solid black) even though subL500 has 96\% of its grid cells empty, compared to MWE. We also add subsubL500 for the NME in the same figure (faint dashed), which corresponds to $n_{\rm particle}=0.0043 h^3{\rm Mpc}^{-3}$ (i.e., 99.6\% of the meshes we used were empty). 
Despite the high sampling noise, it still shows the convergence to the true displacement field in both the power spectrum and the propagator.

For the RREs (Figs.~\ref{figRRE}), $\Lambda=1$ shown in the purple line returns the true nonlinear density field as expected from Eq.~\eqref{deltanrec}, as a sanity check. 
With increasing $\Lambda$, RRE approaches the true displacement field with decreasing BAO damping.
Note that with $\Lambda=10$, the propagator already recovered the propagator of the true displacement.
In Fig.~\ref{figRRE}, the convergence of $\Lambda =100$ and $\Lambda =200$ implies that the correction of $\mathcal O(\Lambda^{-1})$ is negligible for $\Lambda\gtrsim 100$ at least.
However, both curves do not converge to the true displacement field or NME for $k \gtrsim 0.2 \ihMpc$.
As a caveat, we know the initial Lagrangian locations of the particles exactly in this test and therefore ${\bm \chi}=0$ in Eq.~\eqref{noisemod}. 
However, the RREs and true displacement field still disagree.
The disagreement is manifest even for large $\Lambda$, which implies that errors are described by neither $\Lambda$ nor $\bm \chi$.
The discrepancies in L500 should be different from the sampling noise since almost one tracer per mesh is observed in this data.
The solid pink curve in Fig.~\ref{figRRE} shows RRE with $\Lambda =100$ when we assign the rescaled displacement at the {\it center} of the FFT grid where the initial position of each particle falls, rather than at the actual initial particle position. 
This manipulation introduces ${\bm \chi} \ne 0$, but the deviation from the true displacement is removed for L500.
Thus, a part of the unknown error is reduced by this procedure.
In more subsampled cases, the pink and blue dotted curves~(subL500) show a similar noise, which can now be understood as the shot noise.
Thus, centering seems to reduce some of the unknown errors but cannot reduce the sampling noise. 
We note that this empirical observation is missing the theoretical explanation. 
It is not obvious how to apply this `centering' on a highly subsampled case, so the ``centering'' operation is not necessarily useful.
Thus, we conclude that the NME is more robust on scales $k > 0.2 \ihMpc$ than the MWE approach used in Refs.~\cite{Ota:2021caz, Schmittfull:2017uhh}, while the RRE requires further investigation for the error correction.

\section{Post reconstruction estimators}\label{secPost}

So far, we have discussed how to assign given particle displacements to the grid and constructed the displacement field estimators robust to the discreteness effect.
In Figs.~\ref{figNME} and~\ref{figRRE}, we tested our proposed methods for estimating the true displacement field of particles (i.e., in simulations, we knew all initial particle positions a priori and compute the particle displacements by subtracting the initial positions from the final positions). However, this is not the case in real data. We only know the final locations of galaxies in the actual surveys, so we need to reconstruct the displacement field without prior knowledge of the initial mass locations.

\subsection{Displacement field reconstruction}
A displacement field reconstruction method from the observed mass locations is proposed in Ref.~\cite{Schmittfull:2017uhh}.
They proposed to move the mass tracers along the smoothed local density gradient iteratively by progressively reducing the smoothing scale.
After several iterations, they found that the mass configuration becomes almost uniform, and the final locations can be interpreted as estimated Lagrangian positions for each mass.
In this section, we review the mathematical aspect of the algorithm and apply the new estimators NME and RRE on the iteratively reconstructed field.

\medskip
To reconstruct the displacement field from the observed density field and the mass tracer locations, in principle, we have to solve 
\begin{align}
	\delta_{\rm NL}(\mathbf k) &= \int d^3 q e^{-i\mathbf k \cdot \mathbf q} \left(e^{-i\mathbf k \cdot   \mathbf \Psi_{\rm NL}(\mathbf q)}-1 \right)\label{lam1},
\end{align}
for $\mathbf \Psi_{\rm NL}$.
However, this equation is a complicated nonlinear integral equation, which we cannot solve exactly~\footnote{Note that the displacement field divergence is not the log normal field: $\ln[1+\delta_{\rm NL}(\mathbf q)]\neq -\nabla_{\mathbf q}\cdot \mathbf \Psi_{\rm NL}(\mathbf q)$.}.
Let us consider the following eigenfunction decomposition of the perturbation on the flat FLRW background: 
\begin{align}
	\mathbf \Psi_{\rm NL} = i\mathbf k \phi_{\rm NL}+{\bm \beta},
\end{align}
where $\nabla \cdot {\bm \beta}=0$.
Expanding Eq.~\eqref{lam1} to the leading order in $\phi_{\rm NL}$, we get
\begin{align}
	\delta_{\rm NL} \simeq k^2 \phi_{\rm NL}+ R 
	,\label{linear}
\end{align}
where the residual $R$ is the higher order terms in $\phi_{\rm NL}$, and $\bm \beta$ is shown to be small in Ref.~\cite{Baldauf:2015tla}.
The inverse for Eq.~\eqref{linear} is easy if we can ignore $R$.
However, the Zel'dovich approximation $\delta_{\rm NL} \approx k^2 \phi_{\rm NL}$ is valid only for low $k$. 
The iterative reconstruction assumes that there exists a cutoff scale $k_{\rm cut}$ such that $\delta_{\rm NL} \simeq  k^2 \phi_{\rm NL},$ for all $k<k_{\rm cut}$ and then we introduce the smoothed negative displacement
\begin{align}
	\mathbf s = -\frac{i\mathbf k}{k^2}S\delta_{\rm NL},\label{shift}
\end{align} 
where $S=\exp(-k^2/2k^2_{\rm c})$ such that $SR\approx 0$.
This is equivalent to solving the smoothed linearized continuity equation, as we may identify the time derivative of the displacement field with the velocity.  
Then we shift the particles from $\mathbf x$ to $\mathbf x + \mathbf s(\mathbf x)$.
Thus, we partly estimated the Lagrangian position for $k<k_{\rm cut}$ modes.
The new shifted frame would be closer to the uniform Lagrangian frame, so the nonlinearity would be reduced there.
The new cut-off scale can be bigger; $k_{\rm cut, new}>k_{\rm cut}$.
Then we re-estimate the negative displacement for the reduced density perturbations by reducing the smoothing scale as $k_{\rm cut}\to k_{\rm cut, new} = \epsilon k_{\rm cut}$ with $\epsilon>1$ in Eq.~\eqref{shift}.
We repeat this cycle until we derive almost zero $n$-th step displacement. 
The shifted frame is given recursively as  
\begin{align}
	\mathbf x^{(n+1)} = \mathbf x^{(n)} + \mathbf s^{(n)}(\mathbf x^{(n)}),
\end{align}
and, the final location $\mathbf x^{(\infty)}$ will be close to the uniform density frame, which is the estimated Lagrangian position.
In this way, we find a ``coordinate transformation'' from the Eulerian to Lagrangian coordinate in the simulation.
In the iteration, Eq.~\eqref{linear} is generalized to 
\begin{align}
	\delta^{(n)}_{\rm NL} = k^2 \phi_{\rm NL}^{(n)}+ R^{(n)} 
	,\label{linearn}
\end{align}
where the superscript $n$ implies a step of iteration, that is, $\delta_{\rm NL}^{(n)}$ is the density perturbation measured in the $n$-th step particle distribution, and $\phi_{\rm NL}^{(n)}$ is the $n$-th step displacement field.
We attempt to reduce the nonlinearity in the input mass distribution by solving the linear algebra iteratively.
Refs.~\cite{Schmittfull:2017uhh, Ota:2021caz, Seo:2021nev} confirmed that the BAO damping is significantly reduced by reconstructing the displacement field in this way.

Our new estimators reduce the potential discreteness effect in evaluating the reconstructed displacement field. We follow the same iterative reconstruction process as in Ref.~\cite{Ota:2021caz}. The location of the mass tracers after each iteration is identical among the estimators we are testing in this paper.

\subsection{Discrepancy between true and reconstructed displacement fields}\label{sec:distrusim}

\begin{figure*}
	  \centering \includegraphics[width=\linewidth]{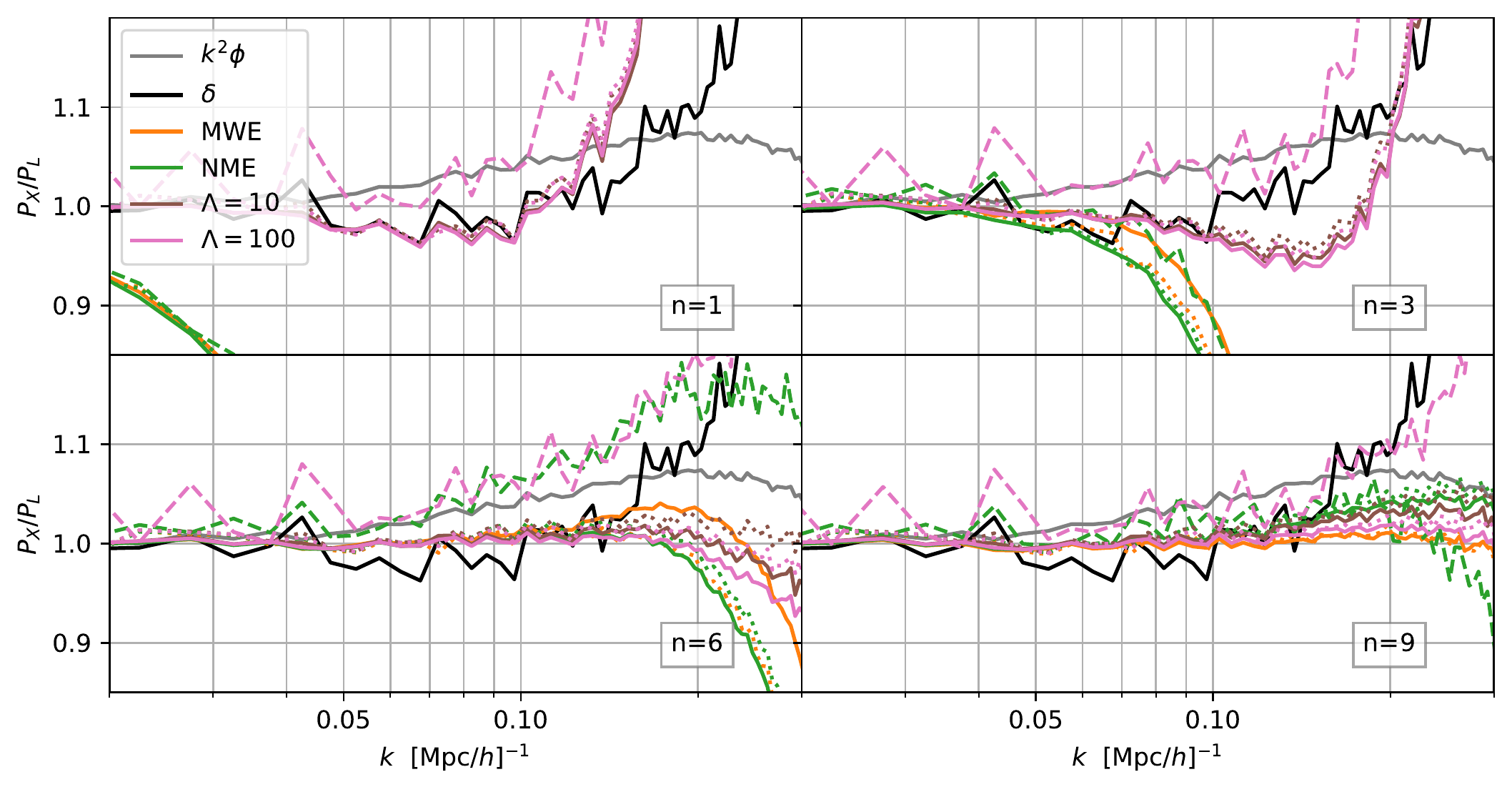}
	  \centering \includegraphics[width=\linewidth]{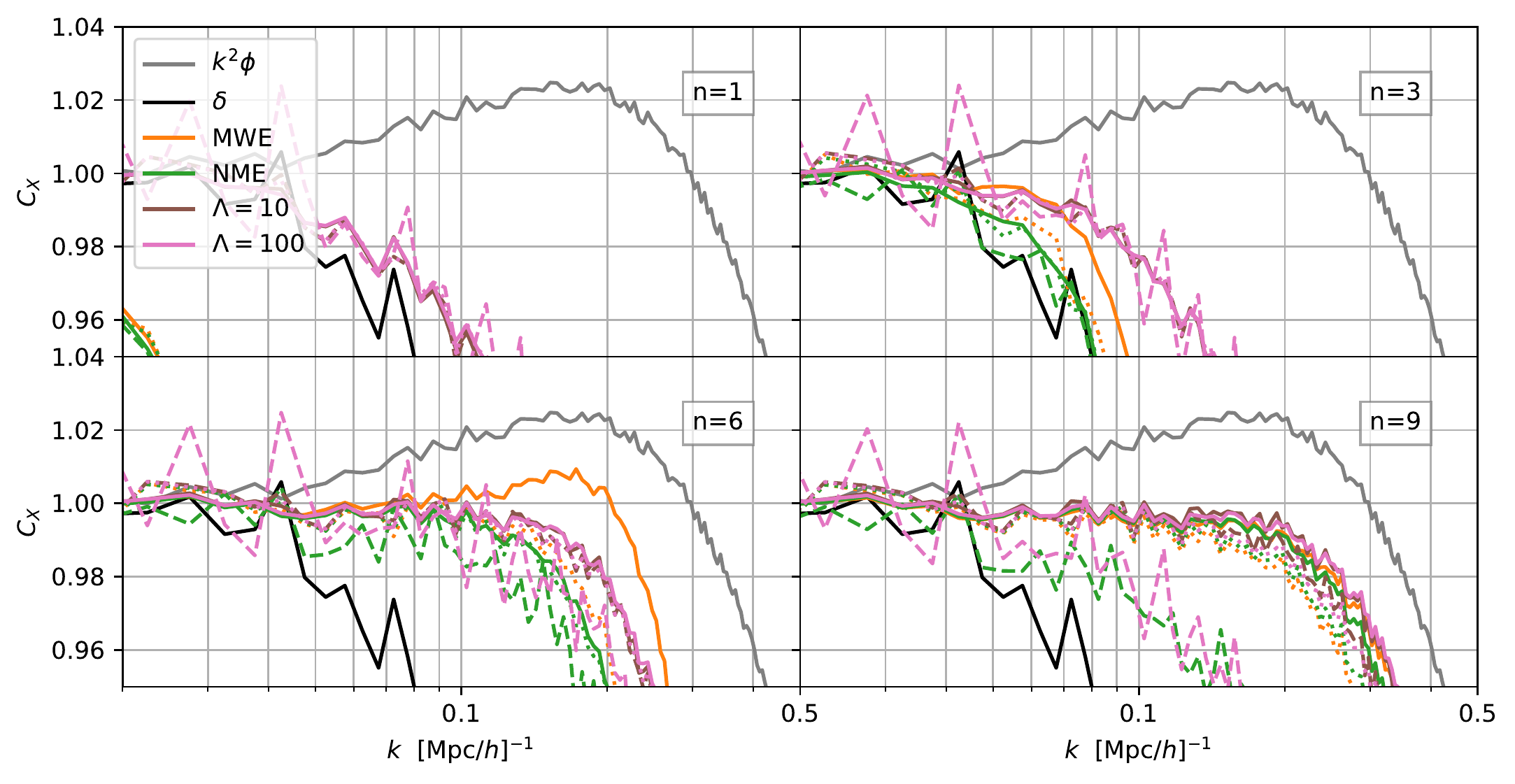}
  \caption{A comparison of the power spectra (top) and propagators (bottom) of the various estimators for iterative reconstruction for $n=1,3,6$, and $9$, in comparison to the true displacement (gray).  The black curve means the correlation functions with the nonlinear density perturbations. A solid~(dotted) curve represents L500~(subL500). 
  Various estimators agree with each other for $n\leq 6$ and $k <0.2\ihMpc$, except for the traditional mass weighted estimators~(orange). The behavior of NME and RRE as a function of $n$ shows a stable trend compared to MWE.
  The dashed lines show subsubL500, which corresponds to $n_{\rm particle}=0.0043 h^3{\rm Mpc}^{-3}$, i.e., closer to a realistic galaxy sample; here, the difference is mainly due to the effect of the sampling noise on the reconstruction efficiency. 
 }
  	\label{itrP}
\end{figure*}

In our earlier work, we also investigated to what extent the iterative procedure reconstructs the true displacement field in the broadband shape of the power spectrum~\cite{Ota:2021caz}.
In that work, we saw about 8\% percent discrepancies between the true displacement field measured in the same simulation and the reconstructed displacement field measured with the MWE estimator for $k >0.1\ihMpc$.
We suspected the inconsistency was partly due to the discreteness effect in the MWE estimator.
In Fig.~\ref{itrP}, we compare the MWE, NME, and RRE estimators of the reconstructed displacement field to revisit this inconsistency since we expect the new estimators are more robust against the discreteness effect. The solid lines are for L500, and the dotted lines are for subL500. Although we expect some sampling noise of the observed field \cite{White2010:1004.0250v1}, MWE shows a drastic difference between the two subsampling rather even though the shot noise of subL500 is still quite negligible, as pointed out in Ref.~\cite{Ota:2021caz}. Our new estimators, NME and RRE are more stable for subsampling than MWE is. 
As we progress to $n>6$, we begin to see differences in the power spectrum on small scales due to the noise of the field affecting reconstruction and also the difference between RRE and NME; because the power spectrum is divided by the linear matter power spectrum, a small residual shot noise will appear significant in this plot. On the other hand, the propagator shows the expected consistency between this two subsampling when using NME and RRE. 
The dashed lines show subsubL500, which corresponds to $n_{\rm particle}=0.0043 h^3{\rm Mpc}^{-3}$, i.e., closer to a realistic dense galaxy sample such as from the Bright Galaxy Survey in Dark Energy Spectroscopic Instrument \cite{DESI:2016fyo}. The reconstruction efficiency of this subsample is expected to be noticeably lower than L500 and subL500, and the small scale will be dominated by shot noise (a typical particle separation of $6.15\hMpc$). Indeed, the maximum efficiency for this sample happens at $n \sim 5- 6$, where the smoothing scale of the step ($3.5-6\hMpc$) is close to this average particle spacing, and a further iteration with a smaller smoothing scale decreases the efficiency.

While robustness against sampling/discreteness is improved, we still found inconsistency with respect to the true displacement field (gray), even on the quasi-linear scale, $k\sim 0.1\ihMpc$, where the estimators are converging towards each other. 
The difference between NME and RRE is small at that scale, so we can now interpret the discrepancy with respect to the true displacement field as an indication that the iterative displacement field reconstruction cannot recover the true displacement field even on the quasi-linear scale perfectly, contrary to our theory expectation in Ref.~\cite{Ota:2021caz}. 
We still recover the initial density information. The propagator in the lower panel of the figure shows that iterative reconstruction is an efficient density field reconstruction method, but it does not fully recover the nonlinear displacement field in the broadband.

Based on this result we argue that the iterative reconstruction method in Ref.~\cite{Schmittfull:2017uhh} does not reproduce the true displacement field. 
Introducing 2LPT or higher-order corrections for Eq.~\eqref{linear} could improve the agreement with respect to the true displacement. Ref.~\cite{Seo:2009fp} implemented 2LPT as an extension to the standard reconstruction, but found a minor improvement in the reconstructed density field. In the current case, the combination of the iterative steps and the 2LPT focusing on the displacement field is worth investigating, since the iterative process extends to the higher $k$ than the standard BAO reconstruction. We leave this investigation for future work.

\subsection{Discrepancy between reconstructed displacement fields and the model}

\begin{figure}
	  \centering \includegraphics[width=\linewidth]{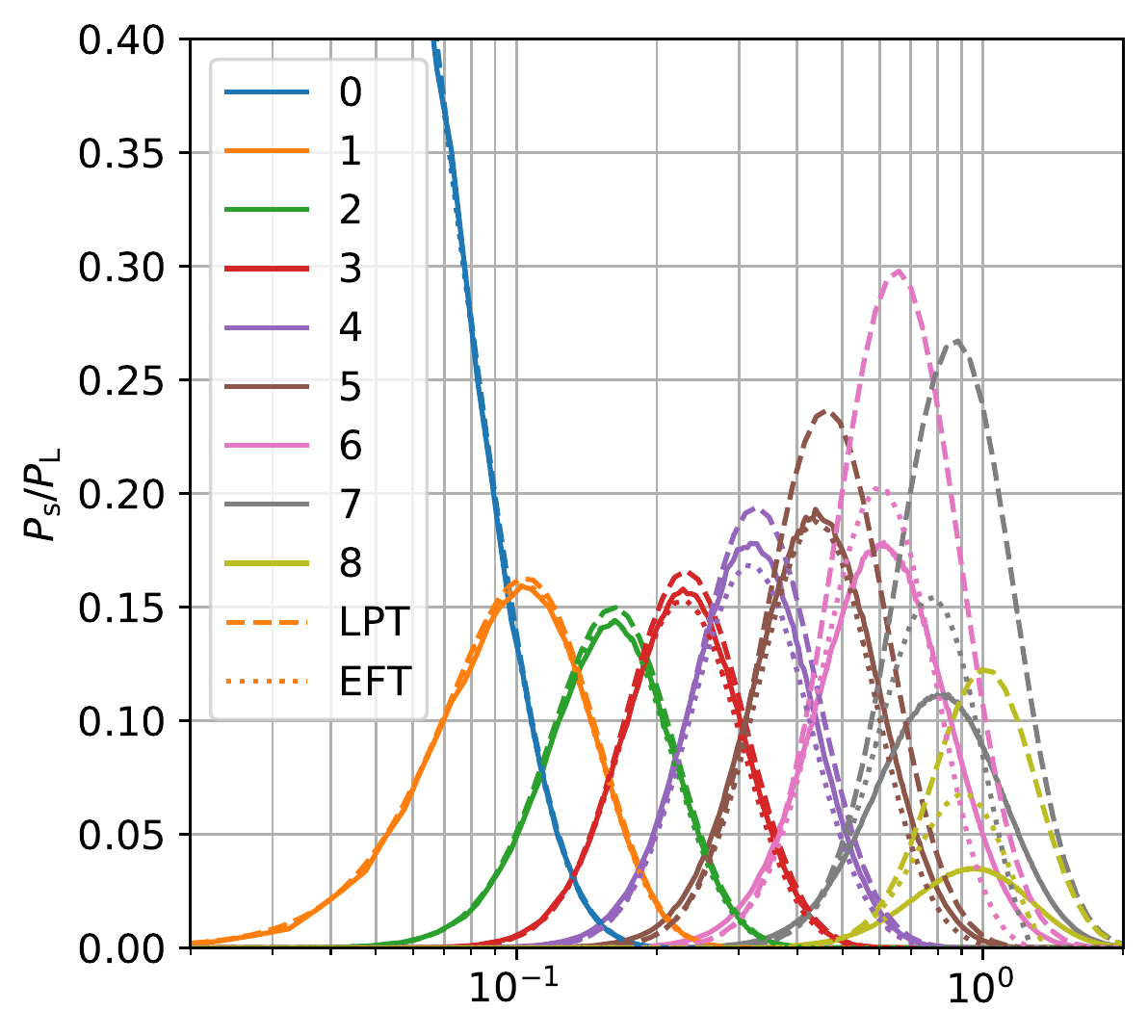}
	  \caption{A comparison of the shift vector (i.e., the input to reconstruction as in Eq. \eqref{shiftn} or \eqref{eftnshift}) power spectra normalized by the linear power spectrum for the simulation (solid), in comparison to our theoretical model based on LPT (dashed), and EFT (dotted) at each iteration.
	  The $y$ axis is the power spectrum of the shift vector normalized by the linear power spectrum, which is the smoothed density fluctuation power spectrum normalized by the linear spectrum.
	  The shift vector of the simulation is better approximated by introducing the EFT terms (in Eq.~\eqref{eftnshift}) up to $n\leq 6$ for $k\leq 0.4\ihMpc$.}\label{dd}
\end{figure}

So far, we have developed methods to reduce the discreteness effect in estimating the displacement fields, which was considered an obstacle for comparing the reconstructed field and the true displacement field in Ref~\cite{Ota:2021caz}. 
Then we found that the discreteness effect cannot explain the discrepancy of 8\% at $k \sim 0.2 \ihMpc$. Note that the discrepancy is apparent even from the very first iteration. Therefore, we revisit and improve the theoretical modeling to address this discrepancy better. 
In Ref.~\cite{Ota:2021caz}, we modeled the reconstruction procedure up to 1-loop order in LPT~(see Appendix~\ref{lartmodeling} for a summary).
A possible cause of discrepancy is that the SPT prediction fails at the quasi-nonlinear scale, i.e. we have $\delta^{(n)}_{\rm NL,SPT} \gg \delta^{(n)}_{\rm NL,sim}$ for $k \gtrsim 0.1 \ihMpc$ at $z=0.6$, and therefore theory overestimates the displacement field~(see Fig.~\ref{dd}).  
As a result, the estimation of the shift vector within SPT
\begin{align}
	\mathbf s^{(n)} \approx -\frac{i\mathbf k}{k^2}S^{(n)}\delta^{(n)}_{\rm NL,SPT},\label{shiftn}
\end{align}
may not be very accurate, depending on the smoothing kernel and the redshift. Indeed Fig.~\ref{dd} shows the difference between Eq.\eqref{shiftn} (dashed curves) and the actual shift-vector (solid curves) from the simulations.
The shortcoming of the 1-loop calculation is mainly because of the UV-sensitive loop integrals.
We can try to correct this term with effective field theory~\cite{Carrasco:2012cv}.
At 1-loop order, a possible EFT correction is written as
\begin{align}
	\mathbf s^{(n)} \approx -\frac{i\mathbf k}{k^2}S^{(n)} \left [ \delta^{(n)}_{\rm NL,SPT} + \alpha k^2  \delta_{\rm L}^{(n)}\right].\label{eftnshift}
\end{align}
We observe the nonlinear density field at every step and find a single parameter $\alpha$ by fitting all steps simultaneously. 
With the least-squares method, we found $\alpha=-0.999\pm 0.011$ by fitting simulations in $0.2<k\,{\rm Mpc}/h<2$~\footnote{$\alpha\sim -1$ is just by accident as $\alpha$ has dimension.}. 
In Fig.~\ref{dd}, we show a comparison of the power spectra of the shift vector for the simulation (solid), LPT (dashed), and EFT (dotted lines). The figure shows that this single EFT term can reduce the discrepancy in $\mathbf s^{(n)}$ between the theory and the simulation, especially for $n\leq 5$. 

\begin{figure*}
\includegraphics[width=\linewidth]{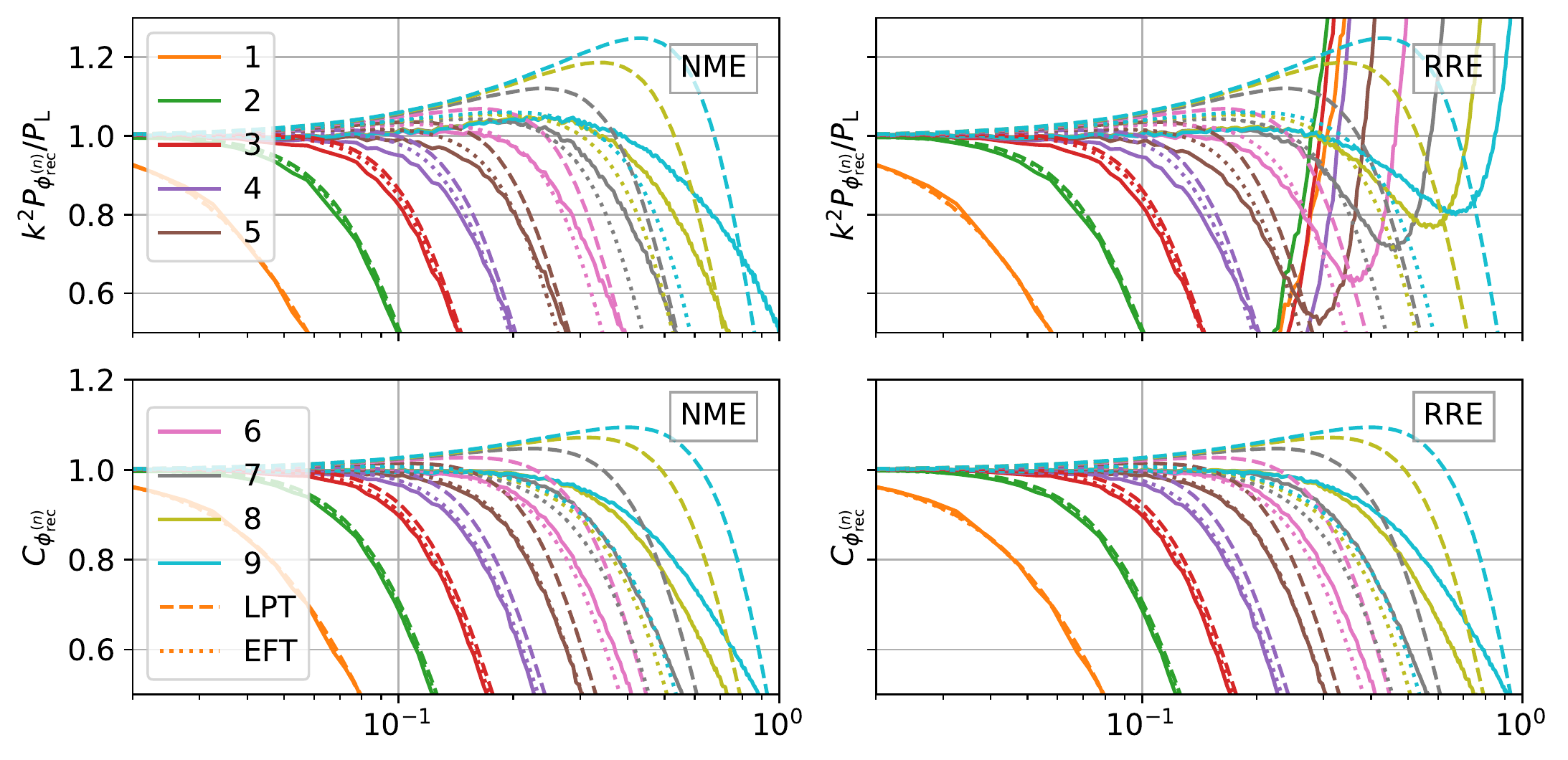}~
  \caption{The post-reconstruction power spectrum normalized by the linear power spectrum (top) and correlation function (bottom) for the simulation, 1-loop LPT, and 1-loop EFT. The EFT parameter is obtained by fitting the shift vector power spectra with 1-loop EFT correction. The simulation estimators are the NME~(left) and the corrected RRE~(right) for L500.
  }
  	\label{itrdata}
\end{figure*}

We now propagate this EFT correction to the reconstructed displacement potential, which is written as 
\begin{align}
	\phi^{(n)}_{\rm rec., EFT} =\phi_{\rm rec.,LPT}^{(n)} - \alpha \delta_{\rm L} \sum_{i=0}^{n-1}  S^{(i)}B^{(i)}.
\end{align}
$B^{(n)}$ is the reconstruction kernel defined in Appendix~\ref{lartmodeling}.
The corresponding post-reconstruction power spectrum is written as 
\begin{align}
	P^{(n)}_{\rm rec,EFT} = P^{(n)}_{\rm rec,LPT} - 2 k^2 \alpha \bar B^{(n)} P_{\rm L}\sum_{i=0}^{n-1}  S^{(i)}B^{(i)},
\end{align}
where $\bar B^{(n)}\equiv 1 - B^{(n)}$.
Fig.~\ref{itrdata} shows a comparison of 1-loop LPT and EFT for iterative reconstruction modeling, and we show our EFT helps to reach a few percent agreements for the displacement field up to $k <0.5\ihMpc$ for $z=0.6$, particularly for NME and for $n <6$ while the agreement is less for RRE. The EFT term can improve the agreement between the assumption in the theoretical model and the simulation and, therefore, the theoretical model of the resulting reconstructed displacement field within 1\% for NME at $k< 0.2\ihMpc$. Therefore our new theory model explains that we do not recover the true displacement field with the current reconstruction method.
While we fixed $\alpha$ in the above consideration, we also varied $\alpha$ for each iteration and considered independent fits; however, it did not improve the EFT correction any further.

In Ref.~\cite{Ota:2021caz}, we tested a different option of adopting EFT, i.e., the EFT fit directly to the post-reconstruction spectrum, by introducing the 1-loop EFT term for the displacement field from the LPT perspective, i.e., we considered
\begin{align}
    \bar P^{(n)}_{\rm rec,EFT}=P^{(n)}_{\rm rec,LPT} + 2 \bar \alpha k^2 P^{(n),{\rm lin}}_{\rm rec,LPT},
\end{align}
with the linear power spectrum of $n$-th step displacement field $\phi^{(n)}_{\rm rec,LPT}$, that is, $P^{(n),{\rm lin}}_{\rm rec,LPT}.$
This prescription is based on the EFT for the displacement field in Ref.~\cite{Baldauf:2015tla}.
Such EFT choice was good at reproducing the true displacement field, but it could not model the power spectrum of the post-reconstruction field.
In detail, the offset between the model and the reconstructed simulation happened mainly in the propagator (approximately $P_{13}$), i.e., the simulation returned $P_{13}$ more negative than the theory. In contrast, the $P_{22}$ contribution seemed in a good agreement.

In this paper, we fit the post-reconstruction spectrum at a few percent precision with the proper choice of the EFT terms.
The corrections in the shift vectors at each iteration with a fitting parameter $\alpha$ are considered.
The new model predicts the $P_{13}$ contribution to 1\% for $n \leq 9$ in the lower panel of Fig.~\ref{itrdata}.
The accuracy of the improved theory model is almost at the same level as the 1-loop EFT fit for the pre-reconstruction matter power spectrum.

\medskip
To summarize, once we correct Eq.~\eqref{shiftn} with the EFT, our model explains that Ref.~\cite{Schmittfull:2017uhh} could not reconstruct the nonlinear displacement field. 
This implies that truncation of $R^{(n)}$ in Eq.~\eqref{linearn} causes the discrepancy in Sec.~\ref{sec:distrusim}.
Thus, at least partly, the nonlinear physics prevents us from recovering the nonlinear displacement field.

\section{Conclusions}

The Lagrangian displacement can be useful in large-scale structure analysis since it contains the baryon acoustic oscillation~(BAO) feature like the linear field. 
Ref.~\cite{Schmittfull:2017uhh} proposed a reconstruction method of the displacement field and confirmed the post reconstruction field is highly correlated with the linear field.
In Ref.~\cite{Ota:2021caz}, we compared the reconstructed displacement field and the true one.
We also constructed a theoretical model for the displacement reconstruction for broadband analysis in that work.
Then we found two discrepancies: the difference between the reconstructed displacement and the true one; and the disagreement between the simulations and model.
The former is the level of 8\%, and the latter is 15\% at $k\sim 0.2 \ihMpc$ for $n=9$ at $z=0.6$.
Sources of the discrepancies were speculated, particularly including a numerical artifact due to the discreteness of the sample.

\medskip
This paper worked on mitigating such numerical artifacts by developing new estimators.
The new estimators are robust to sampling, but we still observed the difference between the true displacement and the reconstructed one.
Therefore, we conclude that the method in Ref.~\cite{Schmittfull:2017uhh} does not reproduce the true displacement field.
Based on the new estimators, we identified the source of the discrepancy between the theoretical model and the reconstruction method.
Using the EFT approach, we improved our theoretical model and decreased the discrepancy between the model and simulation to a few \% at $k\sim 0.2\ihMpc$ for $n=9$ at $z=0.6$.
We summarize our results below.  

\medskip
First, the mass-weighted displacement estimator we used in Ref.~\cite{Ota:2021caz} is subject to the error particularly sensitive to the number of empty pixels.
We proposed two new displacement field estimators to overcome the discreteness effect; the normalized momentum estimator~(NME) and the rescaled resumed estimator~(RRE).
The NME is the momentum-like estimator that does not suffer significantly from the empty pixel. 
Then, utilizing the fact that the Lagrangian positions of the tracers are independent of their displacement field, we showed that the momentum correlation function normalized by the mass correlation function could closely return the displacement field. 
This method can avoid the complexity of direct, volume-weighted estimators using tessellation~\cite{Bernardeau:1995en}.
Another new estimator, RRE, is a density field-based estimator of the displacement field we devised.
Based on the Lagrangian resummation theory~\cite{Matsubara:2007wj}, we showed that a rescaled density field power spectrum asymptotes to the displacement divergence power spectrum.
That is, the RRE reduces to the displacement field divergence spectrum without explicit evaluation of the vector field, and thus we could avoid the empty pixel issue. 
We investigated the convergence of two estimators for various sampling cases. We concluded that NME performs much better than the RRE and MWE for dealing with the sampling artifacts.
We confirmed that NME is stable for $k\lesssim 0.2\ihMpc$ at $z\sim 0.6$ even if 99.6\% of pixels are empty.

\medskip
We applied our new methods for the post-reconstruction displacement fields.
Then we identified the residual difference between the theoretical model in Ref.~\cite{Ota:2021caz} and the reconstructed field in our simulation. 
With a physically motivated implementation of effective field theory for iterative reconstruction, we could produce a  more accurate model for the post-reconstructed field. 
The precision of our model for the broadband of the reconstructed power spectrum is a few \% at $k < 0.2 \ihMpc$ at $z=0.6$.

This work only considered dark matter N-body simulation in real space. 
In terms of the numerical operation, we expect that NME and RRE can be straightforwardly extended to galaxy samples in redshift space, which we plan for future investigation.
Also, we plan to extend our perturbation theory model approach for a more realistic setup with galaxy bias and redshift-space distortion, in addition to the shot noise effect.


\begin{acknowledgments}

The authors thank Marcel Schmittfull for providing simulations and valuable discussions.
AO is supported in part by the National Key R\& D Program of China (No. 2021YFC2203100).
AO and H.-J.S. are supported by the U.S.~Department of Energy, Office of Science, Office of High Energy Physics under DE-SC0019091. 
SS acknowledges support for this work from NSF-2219212.
SS is also supported by World Premier International Research Center Initiative (WPI Initiative), MEXT, Japan. 
This project has received funding from the European Research Council (ERC) under the European Union's Horizon 2020 research and innovation program (grant agreement 853291). FB is a University Research Fellow.

\end{acknowledgments}

\appendix

\begin{widetext}
\section{LPT modeling of iterative reconstruction}
\label{lartmodeling}

The main text considers the effective field theory modeling of post-reconstruction displacement based on the 1-loop LPT modeling discussed in Ref.~\cite{Ota:2021caz}.
In this appendix, we summarize the derivation in the reference.
In that work, we started from a general perturbative expansion ansatz of the $n$-th step nonlinear displacement $\phi^{(n)}$:
\begin{align} 
	k^2\phi^{(n)}(\mathbf k) 
	=&B^{(n)}(\mathbf k) \delta_{\rm L}(\mathbf k)  
	+
	\frac{1}{2!}\int\frac{d^3k_1d^3k_2}{(2\pi)^6}(2\pi)^3 \delta^{(3)}_{\rm D}(\mathbf k-\mathbf k_1-\mathbf k_2) 
	  B^{(n)}(\mathbf k_1,\mathbf k_2)\delta_{\rm L}(\mathbf k_1)\delta_{\rm L}(\mathbf k_2) \notag \\
	 +&
	\frac{1}{3!}\int\frac{d^3k_1d^3k_2d^3k_3}{(2\pi)^9}(2\pi)^3 \delta^{(3)}_{\rm D}(\mathbf k-\mathbf k_1-\mathbf k_2-\mathbf k_3)
		 B^{(n)}(\mathbf k_1,\mathbf k_2,\mathbf k_3)\delta_{\rm L}(\mathbf k_1)\delta_{\rm L}(\mathbf k_2)\delta_{\rm L}(\mathbf k_3).\label{eq27}
\end{align}
We determine the LPT kernels $B^{(n)}$ by solving the recurrence relations, which are derived by modeling the steps summarized in Sec.~\ref{secPost}.
In this appendix, we consider the expansion with respect to $\delta_{\rm L}$, while the original expansion was for $\phi^{(0)}$ since we expanded $\phi^{(0)}$ into $\delta_{\rm L}$ in the end.
We confirmed that the final results are unchanged for both conventions. 
Then, the estimated displacement field is the sum of the total negative displacement 
\begin{align}
    \phi^{(n)}_{\rm rec,LPT} \equiv  \sum_{i=0}^{n-1}\left( \phi^{(i)}- \phi^{(i+1)}\right) =\phi^{(0)}- \phi^{(n)}.\label{def:phirec}
\end{align}
The LPT postreconstruction power spectrum of Eq.~\eqref{def:phirec} up to 1-loop order is written as follows: 
\begin{align}
	&P^{{\rm LPT}}_{\phi^{(n)}_{\rm rec}}=	 
	P_{\phi^{(n)}_{\rm rec}11}
	+
	P_{\phi^{(n)}_{\rm rec}22}
		+
	P_{\phi^{(n)}_{\rm rec}13}
 ,\label{1-loop:power}
\end{align}	
where we defined
\begin{align}
    k^4P_{\phi^{(n)}_{\rm rec}11} = & \left(B^{(0)} - B^{(n)}\right)^2 P_{\rm L}, \\
    k^4 P_{\phi^{(n)}_{\rm rec}22} =	&  \frac{k^3}{4\pi^2} \int_0^\infty x^2dx  \int_{-1}^{1} d\mu  
      P_{\rm L}\left(ky\right)P_{\rm L}(kx) \frac{\left( X^{(0)}- X^{(n)}\right)^2}{2},
	 \\
k^4P_{\phi^{(n)}_{\rm rec}13}=	& 
	 \frac{k^3}{4\pi^2} \bar B^{(n)}P_{\rm L} \int_0^\infty x^2dx  \int_{-1}^{1} d\mu 
	  P_{\rm L}(kx) 
	\left(Y^{(0)} - Y^{(n)} \right).
\end{align}
$X^{(n)}$, $Y^{(n)}$ and $Z^{(n)}$ are defined as  
\begin{align}
	X^{(n)}(k,kx,\mu) &\equiv  B^{(n)}(\mathbf k-\mathbf k',\mathbf k'),\\
	Y^{(n)}(k,kx,\mu) &\equiv  B^{(n)}(-\mathbf k,\mathbf k',-\mathbf k'), \\
	Z^{(n)}(k,kx,\mu) &\equiv  B^{(n)}(-\mathbf k,\mathbf k'),
\end{align}
with $|\mathbf k'|=kx$, $(\mathbf k\times \mathbf k')^2/(kk')^2=1-\mu^2$, and $y^2=1-2x\mu + x^2$.
The recurrence relations for $X$, $Y$ and $Z$ are 
	\begin{align}
	X^{(n+1)} =&X^{(n)}
	-S^{(n)}(k)X^{(n)} \notag \\
	&-
	x^{-1}y^{-2} \mu( 1-x\mu) 
	S^{(n)}(k)B^{(n)}(kx) B^{(n)}(ky) 
	\notag \\
	&- x^{-1}y^{-2}  ( x- \mu  ) (1-x\mu)   S^{(n)}(ky)
	B^{(n)}(kx)  B^{(n)}(ky) \notag \\
	&
	- y^{-2} \mu( x- \mu)  S^{(n)}(kx) B^{(n)}(kx)  B^{(n)}(ky),\label{recalpha}
\\
	Y^{(n+1)} = &Y^{(n)}
    -S^{(n)}(k)Y^{(n)}
     \notag \\
    &
      +2 x^{-2}\mu^2  S^{(n)}(k)  B^{(n)}(k)B^{(n)}(kx)^2
	\notag \\
	&
	+  2\mu^2  S^{(n)}(kx) B^{(n)}(k)B^{(n)}(kx)^2
	\notag \\
    &
	-2x^{-1}y^{-2} \mu
	(1-x\mu) S^{(n)}(k) B^{(n)}(kx) Z^{(n)}  
	\notag \\
    &
   - 2\mu(x-\mu)y^{-2} S^{(n)}(kx) B^{(n)}(kx)Z^{(n)}
   \notag \\
	&   -
	2x^{-1} (x-\mu )(1-x\mu)y^{-2} S^{(n)}(ky)  B^{(n)}(kx)Z^{(n)} 
   	\notag \\
	&-
    2y^{-2} x^{-2}(x-\mu)^2 (1-x\mu)^2  S^{(n)}(ky)
	B^{(n)}(k)B^{(n)}(kx)^2 
  ,\label{recbeta}
\\
	Z^{(n+1)} =&Z^{(n)} 
	-S^{(n)}(ky)Z^{(n)} \notag \\
	&- 
	x^{-1}(x-\mu) (1-x\mu)  
	S^{(n)}(ky) B^{(n)}(k) B^{(n)}(kx) 
	\notag \\
	&-\mu  (x-\mu) S^{(n)}(kx) B^{(n)}(k)  B^{(n)}(kx) \notag \\
	&-x^{-1}\mu (1-x\mu) S^{(n)}(k) 
    B^{(n)}(k)  B^{(n)}(kx)
		.\label{recgamma}
\end{align}
Note that the factors of 2 in Eq.~\eqref{recbeta} comes from the $\mu\to -\mu$ symmetry.
The initial conditions for the recurrence relations are given as 
\begin{align}
	X^{(0)} & = \frac{3(1-\mu^2)}{7y^2}, \\
	Y^{(0)} & = \frac{10(1-\mu^2)^2}{21y^2}, \\
	Z^{(0)} & = \frac{3(1-\mu^2)}{7}.
\end{align}
Similarly, we find that the powerspectra of the shift vectors Eq.~\eqref{shiftn} are written as  
\begin{align}
	P_{s^{(n)}} = P_{s^{(n)}11} + P_{s^{(n)}22} + P_{s^{(n)}13},
\end{align}
where we defined
\begin{align}
	k^4P_{s^{(n)}11} &= B^{(n)}(k)^2P_{\rm L}(k) \\
	k^4P_{s^{(n)}22} &=\frac{k^3}{4\pi^2}\int x^2dx\int d\mu P_{{\rm L}}(kx)P_{{\rm L}}(ky)  \frac{1}{2}\left[  X^{(n)} 
	+ 
	\frac{\mu(1-x\mu)}{xy^2}
	B^{(n)}(kx)B^{(n)}(ky) \right]^2
	 \\
	k^4P_{s^{(n)},13}&=\frac{k^3 }{4\pi^2}B^{(n)} P_{\rm L}\int x^2dx\int d\mu P_{\rm L}(kx)  
	\left[
	 \frac{2(1-x\mu)\mu}{xy^2} Z^{(n)} B^{(n)}(kx) 
	+
	Y^{(n)} - \frac{\mu^2}{x^2} B^{(n)}(kx)^2B^{(n)}(k) \right]  
	.
\end{align}	
We confirmed that the above equation coincides with the 1-loop SPT spectrum for $n=0$.
\end{widetext}

\bibliography{bib}{}

\begin{thebibliography}{10}

\bibitem{Ota:2021caz}
Atsuhisa Ota, Hee-Jong Seo, Shun Saito, and Florian Beutler.
\newblock {Modeling Iterative Reconstruction and Displacement Field in the
  Large Scale Structure}.
\newblock 5 2021.
\newblock \href {http://arxiv.org/abs/2106.00146} {\path{arXiv:2106.00146}}.

\bibitem{Schmittfull:2017uhh}
Marcel Schmittfull, Tobias Baldauf, and Matias Zaldarriaga.
\newblock {Iterative Initial Condition Reconstruction}.
\newblock {\em Phys. Rev. D}, 96(2):023505, 2017.
\newblock \href {http://arxiv.org/abs/1704.06634} {\path{arXiv:1704.06634}},
  \href {https://doi.org/10.1103/PhysRevD.96.023505}
  {\path{doi:10.1103/PhysRevD.96.023505}}.

\bibitem{Eisenstein:2006nj}
Daniel~J. Eisenstein, Hee-jong Seo, and Martin~J. White.
\newblock {On the Robustness of the Acoustic Scale in the Low-Redshift
  Clustering of Matter}.
\newblock {\em Astrophys. J.}, 664:660--674, 2007.
\newblock \href {http://arxiv.org/abs/astro-ph/0604361}
  {\path{arXiv:astro-ph/0604361}}, \href {https://doi.org/10.1086/518755}
  {\path{doi:10.1086/518755}}.

\bibitem{Crocce:2007dt}
Martin Crocce and Roman Scoccimarro.
\newblock {Nonlinear Evolution of Baryon Acoustic Oscillations}.
\newblock {\em Phys. Rev. D}, 77:023533, 2008.
\newblock \href {http://arxiv.org/abs/0704.2783} {\path{arXiv:0704.2783}},
  \href {https://doi.org/10.1103/PhysRevD.77.023533}
  {\path{doi:10.1103/PhysRevD.77.023533}}.

\bibitem{Seo:2009fp}
Hee-Jong Seo, Jonathan Eckel, Daniel~J. Eisenstein, Kushal Mehta, Marc
  Metchnik, Nikhil Padmanabhan, Phillip Pinto, Ryuichi Takahashi, Martin White,
  and Xiaoying Xu.
\newblock {High-Precision Predictions for the Acoustic Scale in the Non-Linear
  Regime}.
\newblock {\em Astrophys. J.}, 720:1650--1667, 2010.
\newblock \href {http://arxiv.org/abs/0910.5005} {\path{arXiv:0910.5005}},
  \href {https://doi.org/10.1088/0004-637X/720/2/1650}
  {\path{doi:10.1088/0004-637X/720/2/1650}}.

\bibitem{Baldauf:2015tla}
Tobias Baldauf, Emmanuel Schaan, and Matias Zaldarriaga.
\newblock {On the Reach of Perturbative Descriptions for Dark Matter
  Displacement Fields}.
\newblock {\em JCAP}, 03:017, 2016.
\newblock \href {http://arxiv.org/abs/1505.07098} {\path{arXiv:1505.07098}},
  \href {https://doi.org/10.1088/1475-7516/2016/03/017}
  {\path{doi:10.1088/1475-7516/2016/03/017}}.

\bibitem{Matsubara:2007wj}
Takahiko Matsubara.
\newblock {Resumming Cosmological Perturbations via the Lagrangian Picture:
  One-Loop Results in Real Space and in Redshift Space}.
\newblock {\em Phys. Rev. D}, 77:063530, 2008.
\newblock \href {http://arxiv.org/abs/0711.2521} {\path{arXiv:0711.2521}},
  \href {https://doi.org/10.1103/PhysRevD.77.063530}
  {\path{doi:10.1103/PhysRevD.77.063530}}.

\bibitem{Eisenstein:2006nk}
Daniel~J. Eisenstein, Hee-jong Seo, Edwin Sirko, and David Spergel.
\newblock {Improving Cosmological Distance Measurements by Reconstruction of
  the Baryon Acoustic Peak}.
\newblock {\em Astrophys. J.}, 664:675--679, 2007.
\newblock \href {http://arxiv.org/abs/astro-ph/0604362}
  {\path{arXiv:astro-ph/0604362}}, \href {https://doi.org/10.1086/518712}
  {\path{doi:10.1086/518712}}.

\bibitem{Tassev:2012hu}
Svetlin Tassev and Matias Zaldarriaga.
\newblock {Towards an Optimal Reconstruction of Baryon Oscillations}.
\newblock {\em JCAP}, 10:006, 2012.
\newblock \href {http://arxiv.org/abs/1203.6066} {\path{arXiv:1203.6066}},
  \href {https://doi.org/10.1088/1475-7516/2012/10/006}
  {\path{doi:10.1088/1475-7516/2012/10/006}}.

\bibitem{Wang_2017}
Xin Wang, Hao-Ran Yu, Hong-Ming Zhu, Yu~Yu, Qiaoyin Pan, and Ue-Li Pen.
\newblock Isobaric reconstruction of the baryonic acoustic oscillation.
\newblock {\em The Astrophysical Journal}, 841(2):L29, jun 2017.
\newblock \href {https://doi.org/10.3847/2041-8213/aa738c}
  {\path{doi:10.3847/2041-8213/aa738c}}.

\bibitem{Zhu:2016xyy}
Hong-Ming Zhu, Ue-Li Pen, and Xuelei Chen.
\newblock {Primordial Density and Bao Reconstruction}.
\newblock 9 2016.
\newblock \href {http://arxiv.org/abs/1609.07041} {\path{arXiv:1609.07041}}.

\bibitem{Zhu:2016sjc}
Hong-Ming Zhu, Yu~Yu, Ue-Li Pen, Xuelei Chen, and Hao-Ran Yu.
\newblock {Nonlinear Reconstruction}.
\newblock {\em Phys. Rev. D}, 96(12):123502, 2017.
\newblock \href {http://arxiv.org/abs/1611.09638} {\path{arXiv:1611.09638}},
  \href {https://doi.org/10.1103/PhysRevD.96.123502}
  {\path{doi:10.1103/PhysRevD.96.123502}}.

\bibitem{Yu_2017}
Yu~Yu, Hong-Ming Zhu, and Ue-Li Pen.
\newblock Halo nonlinear reconstruction.
\newblock {\em The Astrophysical Journal}, 847(2):110, sep 2017.
\newblock \href {https://doi.org/10.3847/1538-4357/aa89e7}
  {\path{doi:10.3847/1538-4357/aa89e7}}.

\bibitem{Shi_2018}
Yanlong {Shi}, Marius {Cautun}, and Baojiu {Li}.
\newblock {New method for initial density reconstruction}.
\newblock {\em \prd}, 97(2):023505, January 2018.
\newblock \href {http://arxiv.org/abs/1709.06350} {\path{arXiv:1709.06350}},
  \href {https://doi.org/10.1103/PhysRevD.97.023505}
  {\path{doi:10.1103/PhysRevD.97.023505}}.

\bibitem{Hada:2018fde}
Ryuichiro Hada and Daniel~J. Eisenstein.
\newblock {An Iterative Reconstruction of Cosmological Initial Density Fields}.
\newblock {\em Mon. Not. Roy. Astron. Soc.}, 478(2):1866--1874, 2018.
\newblock \href {http://arxiv.org/abs/1804.04738} {\path{arXiv:1804.04738}},
  \href {https://doi.org/10.1093/mnras/sty1203}
  {\path{doi:10.1093/mnras/sty1203}}.

\bibitem{Hada:2018ziy}
Ryuichiro Hada and Daniel~J. Eisenstein.
\newblock {Application of the Iterative Reconstruction to Simulated Galaxy
  Fields}.
\newblock {\em Mon. Not. Roy. Astron. Soc.}, 482(4):5685--5693, 2019.
\newblock \href {http://arxiv.org/abs/1810.05026} {\path{arXiv:1810.05026}},
  \href {https://doi.org/10.1093/mnras/sty3137}
  {\path{doi:10.1093/mnras/sty3137}}.

\bibitem{MaoCNN_2021}
Tian-Xiang {Mao}, Jie {Wang}, Baojiu {Li}, Yan-Chuan {Cai}, Bridget {Falck},
  Mark {Neyrinck}, and Alex {Szalay}.
\newblock {Baryon acoustic oscillations reconstruction using convolutional
  neural networks}.
\newblock {\em \mnras}, 501(1):1499--1510, February 2021.
\newblock \href {http://arxiv.org/abs/2002.10218} {\path{arXiv:2002.10218}},
  \href {https://doi.org/10.1093/mnras/staa3741}
  {\path{doi:10.1093/mnras/staa3741}}.

\bibitem{RecCNN:2022}
Christopher~J. {Shallue} and Daniel~J. {Eisenstein}.
\newblock {Reconstructing Cosmological Initial Conditions from Late-Time
  Structure with Convolutional Neural Networks}.
\newblock {\em arXiv e-prints}, page arXiv:2207.12511, July 2022.
\newblock \href {http://arxiv.org/abs/2207.12511} {\path{arXiv:2207.12511}}.

\bibitem{Seo:2021nev}
Hee-Jong Seo, Atsuhisa Ota, Marcel Schmittfull, Shun Saito, and Florian
  Beutler.
\newblock {Iterative reconstruction excursions for Baryon Acoustic Oscillations
  and beyond}.
\newblock 6 2021.
\newblock \href {http://arxiv.org/abs/2106.00530} {\path{arXiv:2106.00530}},
  \href {https://doi.org/10.1093/mnras/stac082}
  {\path{doi:10.1093/mnras/stac082}}.

\bibitem{1980lssu.book.....P}
P.~J.~E. {Peebles}.
\newblock {\em {The large-scale structure of the universe}}.
\newblock 1980.

\bibitem{Bernardeau:1995en}
Francis Bernardeau and Rien van~de Weygaert.
\newblock {A New Method for Accurate Velocity Statistics Estimation}.
\newblock {\em Mon. Not. Roy. Astron. Soc.}, 279:693, 1996.
\newblock \href {http://arxiv.org/abs/astro-ph/9508125}
  {\path{arXiv:astro-ph/9508125}}, \href
  {https://doi.org/10.1093/mnras/279.2.693}
  {\path{doi:10.1093/mnras/279.2.693}}.

\bibitem{Pueblas:2008uv}
Sebastian Pueblas and Roman Scoccimarro.
\newblock {Generation of Vorticity and Velocity Dispersion by Orbit Crossing}.
\newblock {\em Phys. Rev. D}, 80:043504, 2009.
\newblock \href {http://arxiv.org/abs/0809.4606} {\path{arXiv:0809.4606}},
  \href {https://doi.org/10.1103/PhysRevD.80.043504}
  {\path{doi:10.1103/PhysRevD.80.043504}}.

\bibitem{Colombi:2008qz}
Stephane Colombi, Michal Chodorowski, and Romain Teyssier.
\newblock {Cosmic Velocity--Gravity Relation in Redshift Space}.
\newblock {\em Mon. Not. Roy. Astron. Soc.}, 375:348, 2007.
\newblock \href {http://arxiv.org/abs/0805.1693} {\path{arXiv:0805.1693}},
  \href {https://doi.org/10.1111/j.1365-2966.2006.11330.x}
  {\path{doi:10.1111/j.1365-2966.2006.11330.x}}.

\bibitem{Zheng:2013ora}
Yi~Zheng, Pengjie Zhang, Yipeng Jing, Weipeng Lin, and Jun Pan.
\newblock {Peculiar Velocity Decomposition, Redshift Space Distortion and
  Velocity Reconstruction in Redshift Surveys. II. Dark Matter Velocity
  Statistics}.
\newblock {\em Phys. Rev. D}, 88:103510, 2013.
\newblock \href {http://arxiv.org/abs/1308.0886} {\path{arXiv:1308.0886}},
  \href {https://doi.org/10.1103/PhysRevD.88.103510}
  {\path{doi:10.1103/PhysRevD.88.103510}}.

\bibitem{Zhang:2014hra}
Pengjie Zhang, Yi~Zheng, and Yipeng Jing.
\newblock {Sampling Artifact in Volume Weighted Velocity Measurement. I.
  Theoretical Modeling}.
\newblock {\em Phys. Rev. D}, 91(4):043522, 2015.
\newblock \href {http://arxiv.org/abs/1405.7125} {\path{arXiv:1405.7125}},
  \href {https://doi.org/10.1103/PhysRevD.91.043522}
  {\path{doi:10.1103/PhysRevD.91.043522}}.

\bibitem{Zheng:2014ywa}
Yi~Zheng, Pengjie Zhang, and Yipeng Jing.
\newblock {Sampling Artifact in Volume Weighted Velocity Measurement. II.
  Detection in Simulations and Comparison with Theoretical Modeling}.
\newblock {\em Phys. Rev. D}, 91(4):043523, 2015.
\newblock \href {http://arxiv.org/abs/1409.6809} {\path{arXiv:1409.6809}},
  \href {https://doi.org/10.1103/PhysRevD.91.043523}
  {\path{doi:10.1103/PhysRevD.91.043523}}.

\bibitem{Yu:2015gla}
Yu~Yu, Jun Zhang, Yipeng Jing, and Pengjie Zhang.
\newblock {Kriging Interpolating Cosmic Velocity Field}.
\newblock {\em Phys. Rev. D}, 92(8):083527, 2015.
\newblock \href {http://arxiv.org/abs/1505.06827} {\path{arXiv:1505.06827}},
  \href {https://doi.org/10.1103/PhysRevD.92.083527}
  {\path{doi:10.1103/PhysRevD.92.083527}}.

\bibitem{Yu:2016mzj}
Yu~Yu, Jun Zhang, Yipeng Jing, and Pengjie Zhang.
\newblock {Kriging Interpolating Cosmic Velocity Field. II. Taking Anistropies
  and Multistreaming into Account}.
\newblock {\em Phys. Rev. D}, 95(4):043536, 2017.
\newblock \href {http://arxiv.org/abs/1603.05363} {\path{arXiv:1603.05363}},
  \href {https://doi.org/10.1103/PhysRevD.95.043536}
  {\path{doi:10.1103/PhysRevD.95.043536}}.

\bibitem{Park:2000rc}
Changbom Park.
\newblock {Cosmic Momentum Field and Mass Fluctuation Power Spectrum}.
\newblock {\em Mon. Not. Roy. Astron. Soc.}, 319:573, 2000.
\newblock \href {http://arxiv.org/abs/astro-ph/0012066}
  {\path{arXiv:astro-ph/0012066}}, \href
  {https://doi.org/10.1046/j.1365-8711.2000.03886.x}
  {\path{doi:10.1046/j.1365-8711.2000.03886.x}}.

\bibitem{Park:2005bu}
Chan-Gyung Park and Changbom Park.
\newblock {Power Spectrum of Cosmic Momentum Field Measured from the Sfi Galaxy
  Sample}.
\newblock {\em Astrophys. J.}, 637:1--11, 2006.
\newblock \href {http://arxiv.org/abs/astro-ph/0509740}
  {\path{arXiv:astro-ph/0509740}}, \href {https://doi.org/10.1086/498258}
  {\path{doi:10.1086/498258}}.

\bibitem{Howlett:2019bky}
Cullan Howlett.
\newblock {The redshift-space momentum power spectrum \textendash{} I. Optimal
  estimation from peculiar velocity surveys}.
\newblock {\em Mon. Not. Roy. Astron. Soc.}, 487(4):5209--5234, 2019.
\newblock \href {http://arxiv.org/abs/1906.02875} {\path{arXiv:1906.02875}},
  \href {https://doi.org/10.1093/mnras/stz1403}
  {\path{doi:10.1093/mnras/stz1403}}.

\bibitem{Pan:2020thr}
Jun Pan.
\newblock {Estimating Power Spectrum of Discrete Cosmic Momentum Field with
  Fast Fourier Transform}.
\newblock {\em Res. Astron. Astrophys.}, 20:146, 2020.
\newblock \href {http://arxiv.org/abs/2005.06170} {\path{arXiv:2005.06170}},
  \href {https://doi.org/10.1088/1674-4527/20/9/146}
  {\path{doi:10.1088/1674-4527/20/9/146}}.

\bibitem{Seo:2021}
Hee-Jong Seo, Atsuhisa Ota, Shun Saito, Florian Beutler, and Marcel
  Schmittfull.
\newblock {Iterative reconstruction excursions for Baryon Acoustic Oscillations
  and beyond}.
\newblock \href {http://arxiv.org/abs/2021.*****} {\path{arXiv:2021.*****}}.

\bibitem{Ade:2015xua}
P.A.R. Ade et~al.
\newblock {Planck 2015 Results. Xiii. Cosmological Parameters}.
\newblock {\em Astron. Astrophys.}, 594:A13, 2016.
\newblock \href {http://arxiv.org/abs/1502.01589} {\path{arXiv:1502.01589}},
  \href {https://doi.org/10.1051/0004-6361/201525830}
  {\path{doi:10.1051/0004-6361/201525830}}.

\bibitem{Springel:2000yr}
Volker Springel, Naoki Yoshida, and Simon D.~M. White.
\newblock {Gadget: a Code for Collisionless and Gasdynamical Cosmological
  Simulations}.
\newblock {\em New Astron.}, 6:79, 2001.
\newblock \href {http://arxiv.org/abs/astro-ph/0003162}
  {\path{arXiv:astro-ph/0003162}}, \href
  {https://doi.org/10.1016/S1384-1076(01)00042-2}
  {\path{doi:10.1016/S1384-1076(01)00042-2}}.

\bibitem{Springel:2005mi}
Volker Springel.
\newblock {The Cosmological Simulation Code Gadget-2}.
\newblock {\em Mon. Not. Roy. Astron. Soc.}, 364:1105--1134, 2005.
\newblock \href {http://arxiv.org/abs/astro-ph/0505010}
  {\path{arXiv:astro-ph/0505010}}, \href
  {https://doi.org/10.1111/j.1365-2966.2005.09655.x}
  {\path{doi:10.1111/j.1365-2966.2005.09655.x}}.

\bibitem{2018zndo...1451799F}
Yu~{Feng}, Simeon {Bird}, Lauren {Anderson}, Andreu {Font-Ribera}, and Chris
  {Pedersen}.
\newblock {Mp-Gadget/Mp-Gadget: A Tag For Getting A Doi}, October 2018.
\newblock \href {https://doi.org/10.5281/zenodo.1451799}
  {\path{doi:10.5281/zenodo.1451799}}.

\bibitem{White2010:1004.0250v1}
Martin White.
\newblock Shot noise and reconstruction of the acoustic peak.
\newblock 2010.
\newblock URL: \url{https://arxiv.org/abs/1004.0250}, \href
  {http://arxiv.org/abs/Arxiv:1004.0250v1} {\path{arXiv:Arxiv:1004.0250v1}}.

\bibitem{DESI:2016fyo}
Amir Aghamousa et~al.
\newblock {The DESI Experiment Part I: Science,Targeting, and Survey Design}.
\newblock 10 2016.
\newblock \href {http://arxiv.org/abs/1611.00036} {\path{arXiv:1611.00036}}.

\bibitem{Carrasco:2012cv}
John Joseph~M. Carrasco, Mark~P. Hertzberg, and Leonardo Senatore.
\newblock {The Effective Field Theory of Cosmological Large Scale Structures}.
\newblock {\em JHEP}, 09:082, 2012.
\newblock \href {http://arxiv.org/abs/1206.2926} {\path{arXiv:1206.2926}},
  \href {https://doi.org/10.1007/JHEP09(2012)082}
  {\path{doi:10.1007/JHEP09(2012)082}}.

\end{thebibliography}
\bibliographystyle{unsrturl}

\end{document}